\documentclass[%
reprint,
superscriptaddress,
amsmath,amssymb,
aps,
floatfix,
]{revtex4-2}

\usepackage[english]{babel}
\usepackage[colorlinks=true, urlcolor=blue, citecolor=blue, linkcolor=blue ]{hyperref}
\usepackage{graphicx}
\usepackage[euler]{textgreek}

\def\app#1#2{%
	\mathrel{%
		\setbox0=\hbox{$#1\sim$}%
		\setbox2=\hbox{%
			\rlap{\hbox{$#1\propto$}}%
			\lower1.1\ht0\box0%
		}%
		\raise0.25\ht2\box2%
	}%
}
\def\approxprop{\mathpalette\app\relax}

\newlength{\figw}
\newcommand{\sigr}{\sigma^+}
\newcommand{\sigl}{\sigma^-}

\begin{document}
	
	\title{Picosecond x-ray magnetic circular dichroism spectroscopy at the Fe \emph{L}-edges with a laser-driven plasma source}

	
	
	

	\author{Martin Borchert}
	\author{Dieter Engel}
	\author{Clemens von Korff Schmising}
	\author{Bastian Pfau}
	\affiliation{Max-Born-Institut für Nichtlineare Optik und Kurzzeitspektroskopie, Max-Born-Straße 2A, 12489 Berlin, Germany}
	\author{Stefan Eisebitt}
	\affiliation{Max-Born-Institut für Nichtlineare Optik und Kurzzeitspektroskopie, Max-Born-Straße 2A, 12489 Berlin, Germany}
	\affiliation{Technische Universität Berlin, Institut für Optik und Atomare Physik, 10623 Berlin, Germany}
	\author{Daniel Schick}
	\email{schick@mbi-berlin.de}
	\affiliation{Max-Born-Institut für Nichtlineare Optik und Kurzzeitspektroskopie, Max-Born-Straße 2A, 12489 Berlin, Germany}

	\date{\today}
	
	\begin{abstract}
		Time-resolved x-ray magnetic circular dichroism (XMCD) enables a unique spectroscopic view on complex spin and charge dynamics in multi-elemental magnetic materials.
		So far, its application in the soft-x-ray range has been limited to synchrotron-radiation sources and free-electron lasers.
		By combining a laser-driven plasma source with a magnetic thin-film polarizer, we generate circularly polarized photons in the soft x-ray regime, enabling the first XMCD spectroscopy at the Fe $L$ edges in a laser laboratory.
		Our approach can be readily adapted to other transition metal $L$ and rare earth $M$ absorption edges and with a temporal resolution of $<$~10\,ps, a wide range of ultrafast magnetization studies can be realized.
	\end{abstract}
	
	\maketitle
	
	\section{Introduction}
	
	X-ray magnetic circular dichroism (XMCD)~\cite{VanderLaan1986, Schutz1987, Stohr1999} is an established and powerful tool to probe magnetic properties in a variety of material systems, relevant to solid-state physics, chemistry, biology, and material science~\cite{VanderLaan2014}.
	It relies on the strong dichroic absorption of circularly polarized light when tuned close to the vicinity of spin-split core-to-valence-band transitions spanning from the extreme ultraviolet (XUV) up to the hard x-ray regime.
	Especially in the soft x-ray range, the XMCD effect exhibits a large magnetic contrast across the transition metal (TM) $L_{2,3} (2p \rightarrow 3d)$ and rare earth (RE) $M_{4,5} (3d \rightarrow 4f)$ absorption edges~\cite{Kortright1999}.
	In addition to its element- and site-specificity, soft XMCD spectroscopy allows to separate the orbital and spin contributions of the local magnetic moments via so-called sum rules~\cite{Thole1992, Carra1993, Stohr1995}.
	The absorption-based XMCD contrast has enabled further experimental x-ray techniques, e.g. magnetization-sensitive microscopy~\cite{Stohr1993} and holography~\cite{Eisebitt2004}.
	
	Today, circularly polarized soft x-rays are readily available at synchrotron-radiation facilities worldwide, providing a high degree of polarization, high photon flux, and wavelength tunability with superb spectral resolution for tackling a variety of experimental tasks.
	For the ultrafast magnetism community, the availability of short-pulse sources (femto- to few picosecond pulse duration) with inherent circular polarization in the soft x-ray range is, however, rather limited to one free-electron laser~\cite{Higley2019} (FEL) and one laser-slicing source~\cite{Holldack2014}.
	Laser-driven higher-harmonic generation (HHG) sources have gained in popularity and importance for time-resolved XMCD studies~\cite{Siegrist2019, Willems2020, hofherr_ultrafast_2020}, generating circular polarization by two-color driving fields~\cite{Kfir2015, Fan2015} or phase shifters~\cite{Vodungbo2011, Yao2020}.
	However, due to the extremely low flux of these sources~\cite{Feng2020a} at photon energies corresponding to the TM $L$-edges and above, a true laboratory-based alternative for time-resolved XMCD experiments in the soft x-ray range is still lacking.
	
	In this Letter, we describe the first table-top XMCD experiment across the Fe $L$ edges around 700\,eV  with $<$~10\,ps temporal resolution, based on a laboratory source.
	Specifically, the broad-band (50--1500\,eV) emission of a laser-driven plasma x-ray source (PXS) is exploited~\cite{schick_laser-driven_2021}.
	For that, we create a sufficiently large net degree of circular polarization of the -- initially \emph{randomly polarized} -- soft x-rays by the dichroic absorption through a ferrimagnetic thin-film polarizer~\cite{Kortright1997,Pfau2010} with out-of-plane (OOP) magnetization.
	The broadband characteristics of this approach enable the measurement of XMCD spectra across the entire spin-orbit pair of Fe $L$ edges in a single acquisition with a signal-to-noise ratio (SNR) comparable to typical bending-magnet beamlines at synchrotron-radiation facilities.
	In addition, the short-pulse characteristics and the possibility for extension to other photon energies, such as the RE $M$ edges, render this approach a versatile alternative for static and time-resolved XMCD experiments to study a large set of phenomena in ultrafast magnetism, such as all-optical magnetic switching~\cite{Stanciu2007} or photo-induced phase transitions~\cite{Li2022}.
	
	\section{Concept}
	
	\begin{figure*}[htbp!]
		\centering
		\includegraphics[width=0.95\linewidth]{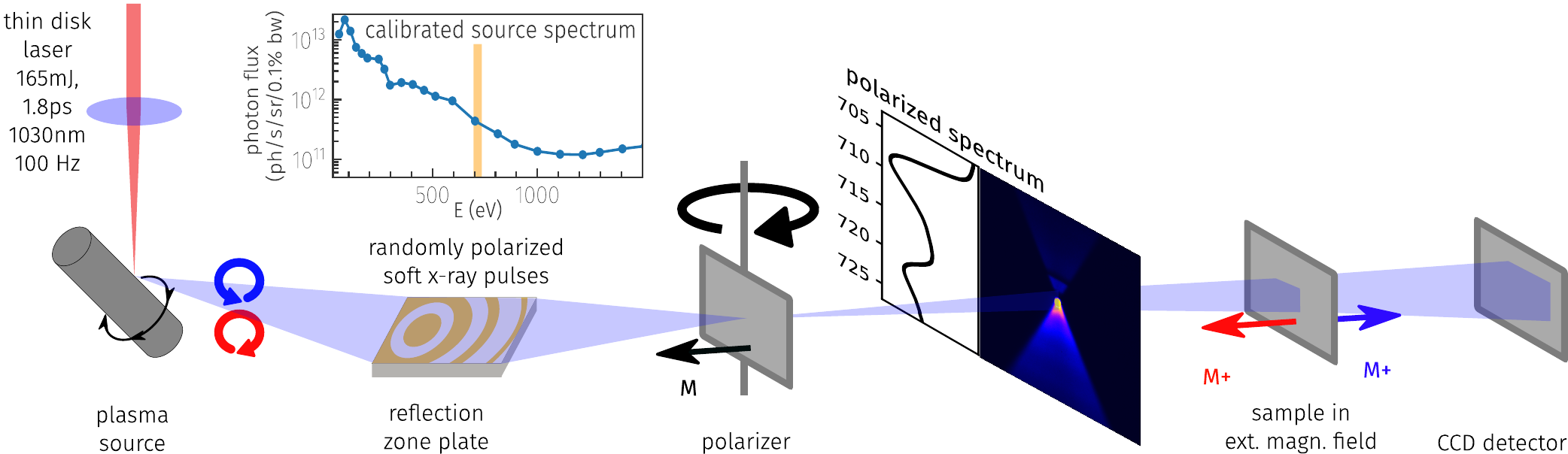}
		\caption{Schematic overview of the experimental setup. Short and intense laser pulses (1.8\,ps, 165\,mJ) are focused onto a spinning tungsten cylinder, generating a plasma, which emits short, broadband soft x-ray pulses ($<$~10\,ps, 50--1500\,eV), see inset.
			A slice of the spectrum is captured by a reflection zone plate and focused through a magnetic thin film polarizer before being transmitted through the sample and detected by a CCD camera.
			The orientation of the polarizer and its magnetization can be changed by a rotational stage, while the magnetization of the sample is altered by an external electromagnet.
		}
		\label{fig:setup}
	\end{figure*}
	
	The driver for the soft-x-ray source is an in-house-developed double-stage thin-disk-amplified laser, producing pulses with a duration of 1.8\,ps full width at half maximum (FWHM) at a wavelength of 1030\,nm with a pulse energy of 165\,mJ at 100\,Hz repetition rate~\cite{Jung2016}.
	At the laser focus of $\approx$ 15\,\textmu m diameter FWHM, a spinning tungsten cylinder is placed under an angle of 45$^\circ$.
	Upon laser-excitation of the tungsten target, a plasma is generated, which emits the aforementioned broadband x-ray radiation into the full solid angle~\cite{mantouvalou_high_2015}.
	We use a reflective zone plate (RZP)~\cite{Brzhezinskaya2013} as the \emph{single} optical element, designed to collect, vertically disperse, and focus the full spectrum across the Fe $L_3$ and $L_2$ edges from the point-like laser-plasma source with a diameter of $\approx$ 40\,\textmu m FWHM through the polarizer and subsequent sample as depicted in Fig.~\ref{fig:setup}.
	The setup provides a photon flux of $10^7$\,ph/s/eV at 700\,eV with a spectral resolution of $E/\Delta E \approx 550$.
	
	We describe the emission from the PXS in the basis of left~($\sigl$) and right~($\sigr$) circularly polarized photons~\cite{Dirac1930}.
	Due to the \emph{thermal} processes involved in the generation, the amounts of $\sigl$ and $\sigr$ polarized photons are equal~\cite{legall_spatial_2004}. 
	This ratio is also maintained after reflection off the RZP, which exhibits a negligible difference ($<$~1\%) in reflectivity for vertical and horizontal field components.
	When transmitting the $\sigl$ and $\sigr$ x-rays through a ferro- or ferrimagnetic polarizer of thickness $d_\text p$, they will exhibit different transmissions $T^\pm(M_\text p, E)$ due to the XMCD effect
	\begin{equation}
		T_\text p^\pm(M_\text p, E) = \exp{\left[- 2 k d_\text p \left(\beta(E) \pm M_\text p\Delta\beta(E)\right)\right]}\ ,\label{eq:trans}
	\end{equation}
	
	where $\beta(E)$ and $\Delta \beta(E)$ are the photon-energy dependent electronic and magnetic absorptive parts of the polarizer's refractive index, respectively, and $M_\text p$ being the magnitude of the polarizer magnetization, projected onto the wavevector of the x-rays with its magnitude $k$.
	Thus, the transmitted light becomes partially circularly polarized as quantified by the polarization factor
	
	\begin{align}
		P(M_\text p, E) & = \frac{T^-(M_\text p, E)-T^+(M_\text p, E)}{T^-(M_\text p, E)+T^+(M_\text p, E)} \nonumber \\
		& = \tanh{\left[2 k d_\text p M_\text p \Delta\beta_\text p(E)\right]} \ , \label{eq:polfac}
	\end{align}
	which is equivalent to the well-known XMCD asymmetry~\cite{Schutz1987}.
	As evident from Eq.~\ref{eq:polfac} a net circular polarization is only observable at photon energies, where $\Delta\beta(E) \neq 0$ and for a finite magnetization $M_\text p$.
	While the magnitude of $\Delta\beta(E)$ is material-specific, the highest magnitude $M_\text p=\pm 1$ is obtained for saturated ferro- and ferrimagnetic films with out-of-plane magnetization at normal incidence.
	The degree of polarization can be enhanced by increasing the thickness of the polarizer but at the same time the total transmission, c.f. Eq.~\ref{eq:trans}, will drop exponentially.
	In order to define an optimum between the degree of polarization and the transmitted light intensity in dependence on the material thickness, the figure of merit $TP^2$~\cite{Kortright1997} is applied.
	
	We choose the ferrimagnetic Gd$_{0.24}$Fe$_{0.76}$ alloy for the polarizer and sample as it exhibits an OOP magnetization in remanence, as verified by optical Kerr measurements.
	The thicknesses of the polarizer and sample are $d_\text p = 100$\,nm and $d_\text s = 55$\,nm, respectively, both being below the optimum of $TP^2$ at the Fe $L_3$ absorption edge, see supplemental material.
	Both films are seeded and capped with 3\,nm of Ta and grown onto a 50\,nm thick SiN membrane with a clear aperture of 4\,mm $\times$ 4\,mm, allowing to transmit the full spatially dispersed soft x-ray spectrum in a single shot.
	Rotating the polarizer by 180$^\circ$ normal to the beam axis, inverts its magnetization $M_\text p$, with respect to the beam propagation direction, allowing to easily invert the circular polarization of the transmitted x-rays.
	The magnetization of the sample $M_\text s$ can be altered by an external magnetic field and the transmission of the polarizer and sample, dispersed along the vertical coordinate, is captured with a subsequent CCD camera.
	
	From the two transmissions $T^+_\text{tot}(E) = T_\text{tot}(M_\text p, +M_\text s, E)$ and $T^-_\text{tot}(E) = T_\text{tot}(M_\text p, -M_\text s, E)$, of the initially randomly polarized soft x-rays through the polarizer (fixed $M_\text p$) and sample (two sample magnetization states $\pm M_\text s$), we derive (c.f. supplemental material) the total asymmetry as
	
	\begin{align}
		A_\text{tot}(M_\text p, M_\text s, E) & = \frac{T_\text{tot}^-(E) - T_\text{tot}^+(E)}{T_\text{tot}^-(E) + T_\text{tot}^+(E)} \label{eq:asymmetry} \\
		& = \tanh{\left[2k d_\text p M_\text p \Delta\beta_\text p(E)\right]} \tanh{\left[2k d_\text s M_\text s \Delta\beta_\text s(E)\right]} \label{eq:asymmetry_tanh}  \\
		& \approxprop  P(M_\text p, E) \ M_\text s\Delta\beta_\text s(E) \label{eq:asymmetry_prop}.
	\end{align}
	In first approximation, $A_\text{tot}$ scales linearly with the degree of circular polarization $P$ induced by the ferrimagnetic polarizer and the sample magnetization component $M_\text{s}$ parallel to the x-ray wavevector.
	
	\section{Results \& Discussion}
\begin{figure}[tb!]
		\centering
		\includegraphics[width=0.9\linewidth]{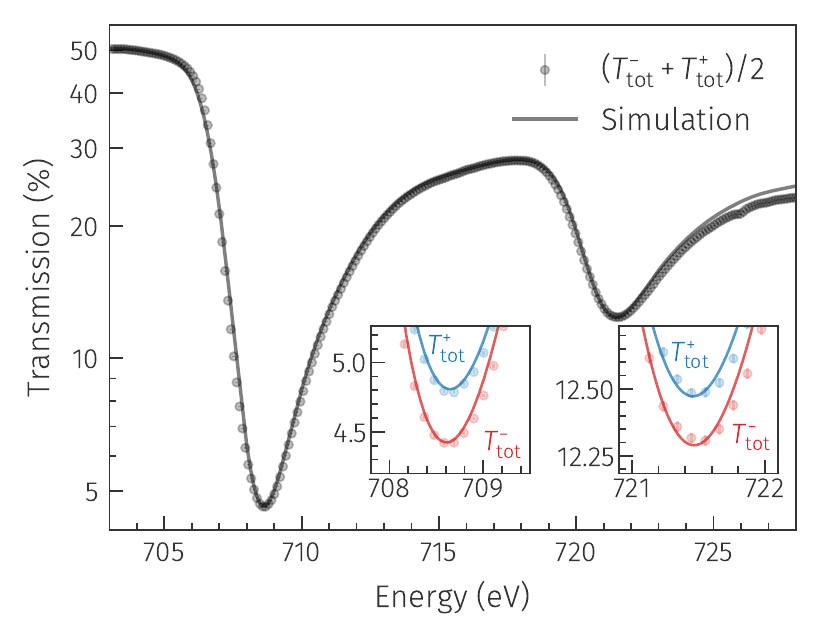}
		\caption{Average of the two total transmissions $T^-_\text{tot}$ and $T^+_\text{tot}$, resembling a typical x-ray-absorption spectrum (XAS).
			The two insets zoom into the Fe $L_3$ (left) and $L_2$ (right) absorption edges and show the individual, transmitted intensities $T^-_\text{tot}$ and $T^+_\text{tot}$ for the two saturated magnetization states of the sample.
			A significant XMCD contrast is visible.
			The solid lines represent magnetic scattering simulations and the dots the experimental data.}
		\label{fig:xmcd_up}
	\end{figure}

	In order to confirm the existence and quantify the degree of the circular polarization of the PXS emission after passing through the ferrimagnetic polarizer of fixed magnetization $M_\text p$, we probe the total, dispersed transmission spectra $T_\text{tot}^\pm$ behind the sample, across the Fe $L_3$ and $L_2$ edges for the two saturation states $M_\text s = \pm 1$ at external magnetic fields of $B = \pm 100\,$mT.
	The average of both dichroic spectra is plotted in Fig.~\ref{fig:xmcd_up} on a logarithmic scale.
	The two insets zoom into the two individual spectra $T_\text{tot}^\pm$ at the $L_3$ and $L_2$ edges, respectively.
	They show a significant difference in absorption between the two magnetic saturation states of the sample -- a clear indication of the XMCD effect and hence of the circular polarization of the impinging x-rays.
	Both spectra have been recorded within an integration time of only 170\,s each, thus less than 6\,min total with an SNR $>$~200 across the whole spectral range.
	
	We quantify the degree of polarization $P(M_\text p, E)$ at the Fe $L_3$ and $L_2$ edges by fitting the experimental spectra with magnetic scattering simulations~\cite{Schick2021b, Elzo2012} based on the sample and polarizer geometries, their experimentally determined $\beta(E)$ and $\Delta\beta(E)$~\cite{Chen1995, Peters2004}, as well as the individual magnetization magnitudes $M_\text{p,s}$ of the polarizer and sample, respectively.
	The simulations have been first bench-marked by XMCD spectra of the sample taken at a synchrotron-radiation beamline, see supplemental material for more details.
	The best fit result, taking the spectral resolution $E/\Delta E$ of the spectrometer into account, yields a maximum degree of circular polarization of 30\,\% at the Fe $L_3$ edge and -7\,\% at the Fe $L_2$ edge, with a magnetization of the polarizer of only $M_\text p = 0.48 <1$.
	We attribute this reduction to magnetically dead layers in the ferrimagnetic Gd$_{24}$Fe$_{76}$ film and/or to magnetic domains having formed, while it was kept in remanence without an external field after being initially saturated once.
	Applying such an external magnetic field by an electromagnet would constantly saturate the polarizer magnetization at $M_\text p=\pm1$, and even allow for fast polarization control.
	
\begin{figure}[tb!]
		\centering
		\includegraphics[width=0.9\linewidth]{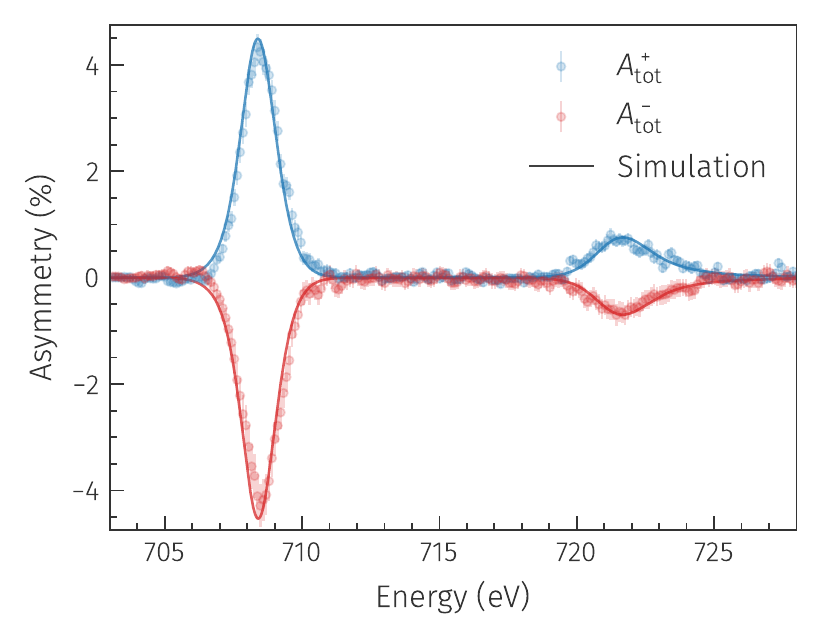}
		\caption{Total asymmetries $A^+_\text{tot} = A_\text{tot} (M_\text p = +1)$ and $A^-_\text{tot} = A_\text{tot} (M_\text p = -1)$ behind the polarizer and sample, calculated from the experimental dichroic spectra from Fig.~\ref{fig:xmcd_up} via Eq.(\ref{eq:asymmetry}).
			The solid lines represent magnetic scattering simulations and the dots the experimental data.
			The colors represent two different polarizer magnetization directions $\pm M_\text p$.
		}
		\label{fig:asym_up_dn}
	\end{figure}

	Further analyzing the same data above, we determine the total asymmetry $A_\text{tot}$, c.f. Eq.~\ref{eq:asymmetry}, and compare it to the simulations, as shown in Fig.~\ref{fig:asym_up_dn}.
	The experimentally observed peak asymmetries at the Fe $L_3$ and $L_2$ edges of about 4\,\% and 1\,\%, respectively, are in good agreement with the theoretical predictions and constitute the first laboratory-based XMCD spectroscopy in the soft-x-ray range.
	The experimental asymmetries are, however, constricted by the non-optimal degree of polarization 
	of the soft x-rays as well as by the limited spectral resolution of the setup.
	Without both of these limitations, the scattering simulations predict even up to 4.5-times larger total asymmetries of 17.5\,\% and 2.5\,\% at the Fe $L_3$ and $L_2$ edges, respectively, for the chosen geometries.
	
	Interestingly, the total asymmetry $A_\text{tot}$ has the same sign at both $L$-edges, although $\Delta\beta(L_3) <$~0 and $\Delta\beta(L_2) >$~0.
	This is best explained by reinspecting Eq.~\ref{eq:asymmetry_tanh}, in which the sign of $\Delta\beta(E)$ determines both the helicity of the x-rays when passing the polarizer as well as the sign of the probed XMCD of the sample at a given photon energy $E$.
	Due to the multiplication of both factors in Eq.~\ref{eq:asymmetry_tanh} the sign of the total asymmetry $A_\text{tot}$ is independent of the sign of $\Delta\beta$.
	It is, however, possible to change the sign of $A_\text{tot}$ by only changing the sign of the polarization factor $P(M_\text p, E)$ at a given photon energy $E$.
	This is accomplished by inverting its magnetization $M_\text p$, i.e. by rotating the polarizer by 180$^\circ$.
	The resulting inverted asymmetry is also plotted in Fig.~\ref{fig:asym_up_dn} and is nearly perfectly symmetric to the initial curve.
	\begin{figure}[tbp!]
		\centering
		\includegraphics[width=0.9\linewidth]{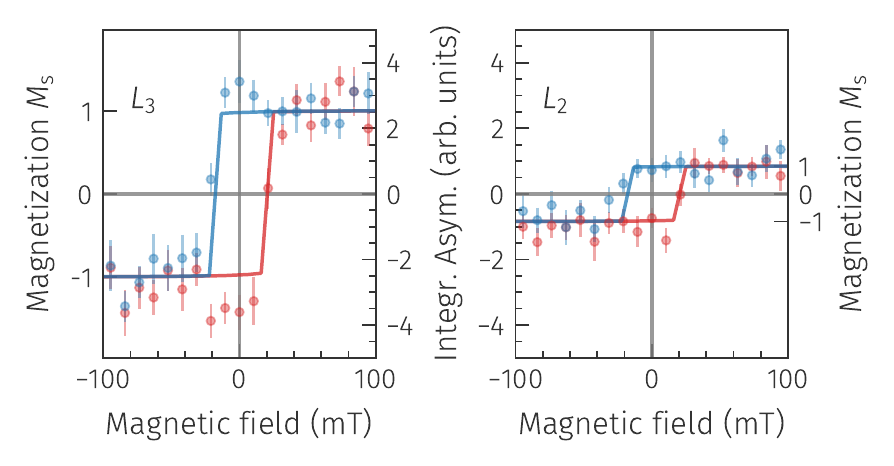}
		\caption{Transmission $T_\text{tot}(M_\text s)$ behind the polarizer and sample at the Fe $L_3$ (left) and $L_2$ (right) absorption edges, while scanning the sample's magnetization $M_\text s$ in a hysteresis loop.
			The inner $y$-axis corresponds to the integrated total asymmetries across the absorption edges.
			The two different colors indicate the two scan directions of the hysteresis loop; the solid lines represent magneto-optical Kerr-effect (MOKE) measurements scaled to the measurement data (dots) as a guide to the eye.
		}
		\label{fig:hyst_up}
	\end{figure}
	
	Finally, to provide an application scenario for our technique, we probe the dispersed, total transmission $T_\text{tot}$ for a fixed polarizer magnetization $M_\text p$, while scanning the sample magnetization via the applied magnetic field from $M_\text s = -1$ to $M_\text s = +1$.
	By plotting the peak intensity, integrated over a 4\,eV bandwidth across each $L$ edge, against the applied magnetic field, we obtain the expected square-shaped magnetic hysteresis loops, as depicted in Fig.~\ref{fig:hyst_up}.
	Magneto-optical Kerr effect (MOKE) measurements from the same sample are overlaid as a guide to the eye.
	We find the expected behavior of the hysteresis scans with a smaller amplitude at the Fe $L_2$ compared to the $L_3$ edge, but with the same symmetry, i.e. observation of positive asymmetry for positive field and vice versa.
	The latter observation can be again explained by the different signs of the polarization factor at both edges.
	Note, that the integration time per data point in the hystereses scans was reduced by an order of magnitude to only 17\,s, compared to the spectra shown in Figures~\ref{fig:xmcd_up} and \ref{fig:asym_up_dn}, and that they are both obtained simultaneously.
	
	The first realization of a picosecond, laser-driven XMCD spectrometer reaching the TM $L$ edges is an important step towards opening this powerful technique for a broader community.
	Next, we want to further discuss the applicability of our approach in terms of available photon flux and SNR, as well as of spectral and temporal resolution, in particular for the field of ultrafast magnetism research.
	
	The SNR $>$~200 of the dichroic spectra across the Fe $L_3$ and $L_2$ edges was achieved within only a few minutes of total integration time.
	This data quality is comparable to current bending-magnet beamlines at synchrotron-radiation facilities, c.f. supplemental material, and is the result of the spectroscopic \emph{white-light} scheme and the possible self-normalization to non-resonant regions, when evaluating these broadband spectra.
	It is hence possible to acquire low-noise asymmetry spectra with partially circularly polarized soft x-rays, which can be fitted against magnetic scattering simulations with very high sensitivity.
	These simulations in combination with a calibrated polarizer even allow for determining the \textit{absolute} XMCD of the investigated sample.
	
	Despite the remarkable SNR of the results presented, the setup can be improved for XMCD spectroscopy in several aspects.
	As already mentioned above, magnetic field control at the polarizer would ensure its full performance and allow for fast polarization switching.
	Even more important, we have identified intensity fluctuations of the PXS as the main source of noise in the experiment, since the detectable photon flux, even at the strongly absorbing Fe $L_3$ edge, is still well above the noise level of the CCD camera.
	These fluctuations mainly originate from mechanical instabilities of the target cylinder and can be tackled by improved suspension and new target concepts, such as metallic tapes or wires, as well as by applying additional normalization schemes, e.g. by a spectrally resolved $I_0$-monitor~\cite{Yao2020}.
	While reducing the noise, it is also possible to increase the signal, namely the XMCD asymmetry, by increasing the polarizer thickness and hence the degree of circular polarization.
	From our simulations, we predict an increase of the maximum degree of circular polarization at the Fe $L_3$ edge for a 200\,nm (400\,nm) thick polarizer of 85\% (99\%) at a total transmission of 1.0\% (0.04\%) for the current 50\,nm sample.
	Although the detectable photon flux is drastically reduced by one to two orders of magnitude for such polarizer thicknesses, it will at the same time allow for single photon counting schemes, using novel area detector generations~\cite{Bergamaschi2018a, Dullin2018}, for which the noise of the experiment is finally approaching the shot noise limit.
	Moreover, the 100\,Hz repetition rate of the experiment is currently determined by the laser system, while the mechanical source and detection would easily allow for repetition rates of $>$1000\,Hz.
	Hence, using commercially available laser systems~\cite{Pfaff2021}, it is feasible to increase the available photon flux by more than one order of magnitude.
	
	While the available photon flux and SNR are certainly well-suited for static experiments, we want to consider the particular value of an XMCD spectrometer with $<$~10\,ps temporal resolution in ultrafast magnetism research, as alternative sources are rare.
	While many of the fundamental processes during laser-driven spin dynamics occur on a sub-picosecond time scale~\cite{Battiato2010, Dewhurst2018, Beens2020}, many collective phenomena such as magnetic phase transitions~\cite{Li2022}, spin precessions~\cite{Burn2020}, and all-optical magnetic switching (AOS)~\cite{Radu2011, Steinbach2022} share intrinsic time scales of a few pico- up to nanoseconds.
	For instance, the realization of magnetic switching exploiting few-picosecond laser- or current pump pulses~\cite{igarashi2020,Yang2017, Grimaldi2020} is of high technological relevance for better CMOS integrability.
	To that end, our setup provides a new laboratory-based tool to specifically follow the sub-lattice magnetization of typical heterostructures and alloys, made of RE and TM metals, during such picosecond magnetization switching dynamics and can provide relevant insights into the underlying mechanisms, which are otherwise hardly accessible, e.g. via optical methods.
	As in such experiments, the magnetic information is usually determined by integrating the XMCD asymmetry over the full magnetically sensitive absorption edges (e.g. the Fe $L_3$ and $L_2$ edges), the moderate spectral resolution of $E/\Delta E \approx 550$ and the photon-energy-dependent degree of circular polarization do not hamper the application of our method.
	
	
	\section{Conclusion}
	So far, all studies on magnetization dynamics in the lower picosecond regime, exploiting the XMCD effect at photon energies above the water window ($>$~500\,eV) require the use of large-scale facilities.
	As an alternative, we have realized the first XMCD spectroscopy setup in the soft-x-ray regime at the Fe $L$-edges at a laser-driven source. 
	Our concept can be readily extended to other photon energies, namely to the magnetically relevant TM $L$ and RE $M$-edges from approx. 500-1500\,eV.
	The high SNR of the presented XMCD spectra in combination with the few-picosecond temporal resolution will enable the study of a wide range of magnetization dynamics relevant for, e.g. all-optical magnetic switching or magneto-structural phase transitions.
	The high flexibility in sample environment, the short iteration cycles, e.g. for sample customization, and the high availability of an laboratory-based ultrafast XMCD setup will allow for detailed and systematic studies of time- and element-resolved spin dynamics.
	
	\textbf{Acknowledgments}
	M.B., C.v.K.S., and S.E. would like to thank the DFG for funding through TRR227 project A02.
	
	\textbf{Author Contributions}
	D.S. conceived the experiment together with M.B., B.P., and S.E.; D.E. grew the samples; M.B. carried out the experiments and evaluated the data; M.B. and D.S. discussed the results and wrote the manuscript with contributing from all other co-authors.
	
	\textbf{Data availability}
	The data sets generated during and/or analyzed during the current study are available from the corresponding author on reasonable request.
	
	\textbf{Competing Interests}
	The authors declare that they have no competing financial or non-financial interests.
	
	\textbf{Supplemental document}
	See Supplement 1 for supporting content.

	
	%

\clearpage

\widetext
\begin{center}
	\textbf{\large Supplemental Materials: Picosecond x-ray magnetic circular dichroism spectroscopy at the Fe \emph{L}-edges with a laser-driven plasma source}
\end{center}
\setcounter{equation}{0}
\setcounter{figure}{0}
\setcounter{table}{0}
\setcounter{page}{1}
\setcounter{section}{0}
\makeatletter
\renewcommand{\theequation}{S\arabic{equation}}
\renewcommand{\thefigure}{S\arabic{figure}}
\renewcommand{\bibnumfmt}[1]{[S#1]}
\renewcommand{\citenumfont}[1]{S#1}
\renewcommand{\thesection}{S-\Roman{section}}

\section{Derivation of the analytical model}

\subsection{One thin film}

The relative transmission $T$ through a non-magnetic sample is

\begin{align}
	T &= e^{-2kd\beta(E)},
	\intertext{where $k = 2\pi/\lambda$ is the magnitude of the wave vector $\vec{k}$.
		Circularly polarized light (CPL) transmits differently if the sample is fully magnetized, depending on the helicity ($\sigma_+$ vs. $\sigma_-$).}
	T^+(M, E) &= e^{-2kd\beta(E)_+} = e^{-2kd(\beta(E) + M\Delta \beta(E))} = e^{-2kd\beta(E)}e^{-2kdM\Delta\beta(E)} \\
	T^-(M, E) &= e^{-2kd\beta(E)_-} = e^{-2kd(\beta(E) - M\Delta \beta(E))} = e^{-2kd\beta(E)}e^{+2kdM\Delta\beta(E)} \\
	\intertext{For elliptically polarized light consisting of $P \sigma_+$ and $(1-P) \sigma_-$ light, the total transmission is}
	T(M,E,P) &= P T^+(M, E) + (1-P) T^-(M, E) \label{eq:generalDichroism}\\
	\intertext{For the special case of linear polarized light (LPL) or randomly polarized light (RPL) consisting of $P$ = 50\% of $\sigma_+$ and $1-P$ = 50\% of $\sigma_-$ light, this simplifies to}
	T(M,E,P=0.5)  &= 0.5 (e^{-2kd\beta(E)}e^{-2kd M \Delta \beta(E)} + e^{-2kd\beta(E)}e^{+2kd M \Delta \beta(E)}) \\
	&= e^{-2kd\beta(E)} \cosh(2kdM\Delta\beta(E))
	\intertext{From this, we can derive the known~\cite{Schutz1987} XMCD asymmetry}
	A_\text{XMCD} &= \frac{T_\text{diff}}{T_\text{sum}}, \\
	\intertext{with}
	T_\text{diff}(M,E,P) &= P T^+(M, E) + (1-P) T^-(M, E) - P T^+(-M, E) - (1-P) T^-(-M, E) \\
	&= Pe^{-2kd\beta(E)}e^{-2kdM\Delta\beta(E)} + (1-P)e^{-2kd\beta(E)}e^{+2kdM\Delta\beta(E)}\\
	&\, - Pe^{-2kd\beta(E)}e^{+2kdM\Delta\beta(E)} - (1-P)e^{-2kd\beta(E)}e^{-2kdM\Delta\beta(E)} \\
	&= 2e^{-2kd\beta(E)}(1-2P)\sinh(2kdM\Delta\beta(E))
	\intertext{and}
	T_\text{sum}(M,E,P) &= P T^+(M, E) + (1-P) T^-(M, E) + P T^+(-M, E) + (1-P) T^-(-M, E) \\
	&= Pe^{-2kd\beta(E)}e^{-2kdM\Delta\beta(E)} + (1-P)e^{-2kd\beta(E)}e^{+2kdM\Delta\beta(E)}\\
	&\, + Pe^{-2kd\beta(E)}e^{+2kdM\Delta\beta(E)} + (1-P)e^{-2kd\beta(E)}e^{-2kdM\Delta\beta(E)} \\
	&= 2e^{-2kd\beta(E)}\cosh(2kdM\Delta\beta(E)) \label{eq:cosh_derivation}
	\intertext{this yields}
	A(M,E,P) &= \frac{T_\text{diff}}{T_\text{sum}} \\
	&= \frac{2e^{-2kd\beta(E)}(1-2P)\sinh(2kdM\Delta\beta(E))}{2e^{-2kd\beta(E)}\cosh(2kdM\Delta\beta(E))} \\
	&= \frac{(1-2P)\sinh(2kdM\Delta\beta(E))}{\cosh(2kdM\Delta\beta(E))} \\
	&= (1-2P) \tanh{(2kdM\Delta\beta(E))}
	\intertext{Thus, for RPL or LPL with $P = 0.5$, $A = 0$, and for CPL with $P = 1$, this simplifies to the known XMCD asymmetry~\cite{Schutz1987}}
	A(M,E,P=1) &= -\tanh{(2kdM\Delta\beta(E))}.
\end{align}

\subsection{Two consecutive films}

For a given magnetization of the polarizer $M_ \text p$ and a magnetization of the sample $M_ \text s$, the total transmission $T_\text{tot}(E)$ for LPL or RPL becomes:

\begin{align}
	T_\text{tot}(M_\text p,M_\text s,E) &= \frac{1}{2}T^0_\text p e^{-2kd_\text p M_\text p\Delta \beta_\text p(E)} T^0_\text s e^{-2kd_\text s M_\text s\Delta \beta_\text p(E)_\text s} \\
	&+ \frac{1}{2}T^0_\text p e^{+2kd_\text p M_\text p\Delta \beta_\text p(E)} T^0_\text s e^{+2kd_\text s M_\text s\Delta \beta_\text p(E)_\text s}.
\end{align}

Here, $T^0_\text{p,s}(E) = \exp(-2kd_\text{p,s}\beta_\text{p,s}(E))$ is the non-magnetic transmission through the polarizer and sample, respectively.
We now compare the two cases $T_\text{tot}^\pm$ where the magnetization of the sample is reversed from $M_\text s = +1$ to $M_\text s = -1$ and the magnetization of the polarizer $M_ \text p$ is fixed.
We calculate the asymmetry $A_\text{tot}$ as the ratio of difference and sum and by using hyperbolic trigonometric identities.

\begin{align}
	A_\text{tot}(M_\text p, M_\text s,E) &=\frac{T_\text{diff}}{T_\text{sum}}\\
	T_\text{sum} &= T_\text{tot}^+ + T_\text{tot}^-\\
	&= T^0_\text p T^0_\text s e^{-2kd_\text p M_\text p\Delta \beta_\text p(E)} \cosh\left(2kd_\text s M_\text s\Delta \beta_\text p(E)\right)\\
	&\,+ T^0_\text p T^0_\text s e^{+2kd_\text p M_\text p\Delta \beta_\text p(E)} \cosh\left(2kd_\text s M_\text s\Delta \beta_\text p(E)\right)\\
	&= 2 T^0_\text p T^0_\text s \cosh\left(2kd_\text p M_\text p\Delta \beta_\text p(E)\right) \cosh\left(2kd_\text s M_\text s\Delta \beta_\text p(E)\right).\\
	T_\text{diff} &= T_\text{tot}^+ - T_\text{tot}^-\\
	&= T^0_\text p T^0_\text s e^{-2kd_\text p M_\text p\Delta \beta_\text p(E)} \sinh\left(2kd_\text s M_\text s\Delta \beta_\text p(E)\right)\\
	&\,-T^0_\text p T^0_\text s e^{+2kd_\text p M_\text p\Delta \beta_\text p(E)} \sinh\left(2kd_\text s M_\text s\Delta \beta_\text p(E)\right)\\
	&= 2 T^0_\text p T^0_\text s \sinh\left(2kd_\text p M_\text p\Delta \beta_\text p(E)\right) \sinh\left(2kd_\text s M_\text s\Delta \beta_\text p(E)\right). \\
	A_\text{tot}(M_\text p, M_\text s,E) &= \tanh\left(2kd_\text p M_\text p\Delta \beta_\text p(E)\right)\tanh\left(2kd_\text s M_\text s\Delta \beta_\text p(E)_\text s\right).
\end{align}

\section{Methods}
The generated soft x-rays at the plasma x-ray source show non-negligible fluctuations in pointing and intensity due to the imperfection of the target cylinder and the stability of the high-intensity driver laser.
As an optimum between long read-out times for the camera and shot-to-shot fluctuations of the source, we integrate 100 laser pulses at the source's repetition rate of 100\,Hz for one acquisition.
Summing the detector counts normal to the dispersion direction (horizontal axis) and subtracting the background results in the full magnetic transmission spectrum $T_\text{tot}(M_\text{p},M_\text{s},E)$ along the dispersion direction (vertical axis).
Before averaging multiple of these spectra, we slightly shift them along the (vertical) photon energy axis for their absorption maxima to match and normalize them to the same value before the resonance to account for both, the pointing and intensity fluctuations.

\section{Simulations}

We model the experimentally acquired transmission results at the plasma x-ray source by magneto-optical simulations via the \textsc{udkm1Dsim} package~\cite{Schick2021b}.
For that, we input the detailed layer structure of the sample and polarizer.
For each material layer, the tabulated value for its density and the nominal values for its thickness (as determined during the growth process) are used.
Only the very sensitive thickness and composition of the magnetic Gd$_{0.24}$Fe$_{0.76}$ layers have been varied in the simulations in order to fit the experimental results. 
For most materials, the atomic scattering factors can be approximated from Chantler et al.~\cite{Chantler2001} and the magnetic scattering factors can be set to zero within the photon energy range of interest (690-750\,eV).
For Fe, we use experimentally determined atomic and magnetic scattering factors from literature~\cite{Chen1995,Peters2004} as they accurately represent the dichroic $L_2$ and $L_3$ absorption resonances.
We add a small linear offset (1.2\% absolute transmission at 703\,eV to 0.2\% at 728\,eV) to the simulated spectra for perfect matching of the experimental result.
This offset possibly stems from the inhomogeneous detector response and optical errors of the RZP.
We have acquired additional XMCD spectra of the Gd$_{0.24}$Fe$_{0.76}$ sample structure at a synchrotron-radiation beamline, c.f. Section~\ref{sec:PM3}, in order to independently benchmark the simulations and to minimize the number of free parameters in the actual fits of the simulations to the data from the laser-based experiments.

\section{Figure of merit for thin film thicknesses}

\begin{figure}[htbp!]
	\centering
	\includegraphics[width=0.75\linewidth]{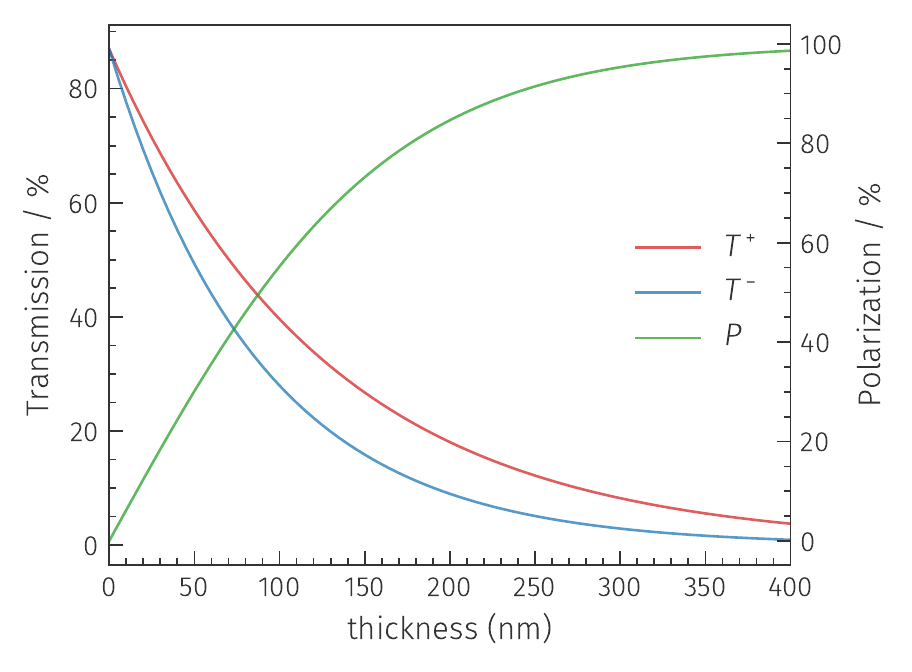}
	\caption{Transmission as a function of Gd$_{0.24}$Fe$_{0.76}$ layer thickness for $\sigma_+$ and $\sigma_-$ light as well as the resulting degree of circular polarization.}
	\label{fig:fom_1}
\end{figure}

To estimate the optimal thickness of the Gd$_{0.24}$Fe$_{0.76}$ layer in the polarizer and sample we calculate the transmission $T^\pm$ for $\sigma_+$ and $\sigma_-$ light as well as the resulting degree of circular polarization $P$ for each structure, see Fig.~\ref{fig:fom_1}.

\begin{figure}[htbp!]
	\centering
	\includegraphics[width=0.75\linewidth]{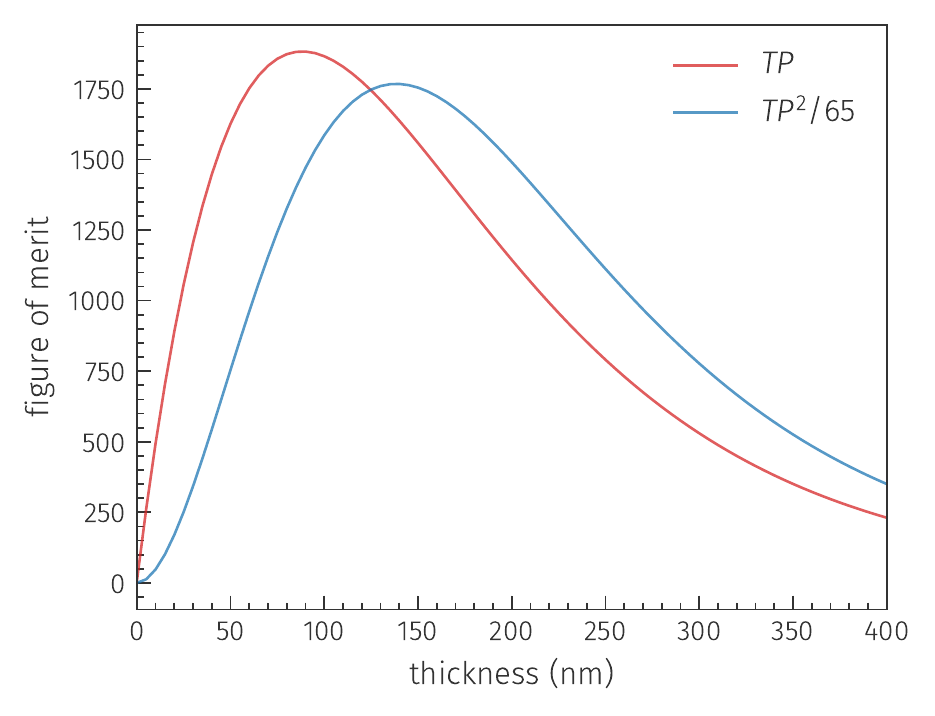}
	\caption{{Figure of merit $TP$ and $TP^2$ as a function of Gd$_{0.24}$Fe$_{0.76}$ layer thickness.}}
	\label{fig:fom_2}
\end{figure}

From that we can calculate figures of merit $TP^2$ as shown in Fig.~\ref{fig:fom_2}.
For comparison we also show an alternative figure of merit $TP$.

\begin{figure}[htbp!]
	\centering
	\includegraphics[width=0.75\linewidth]{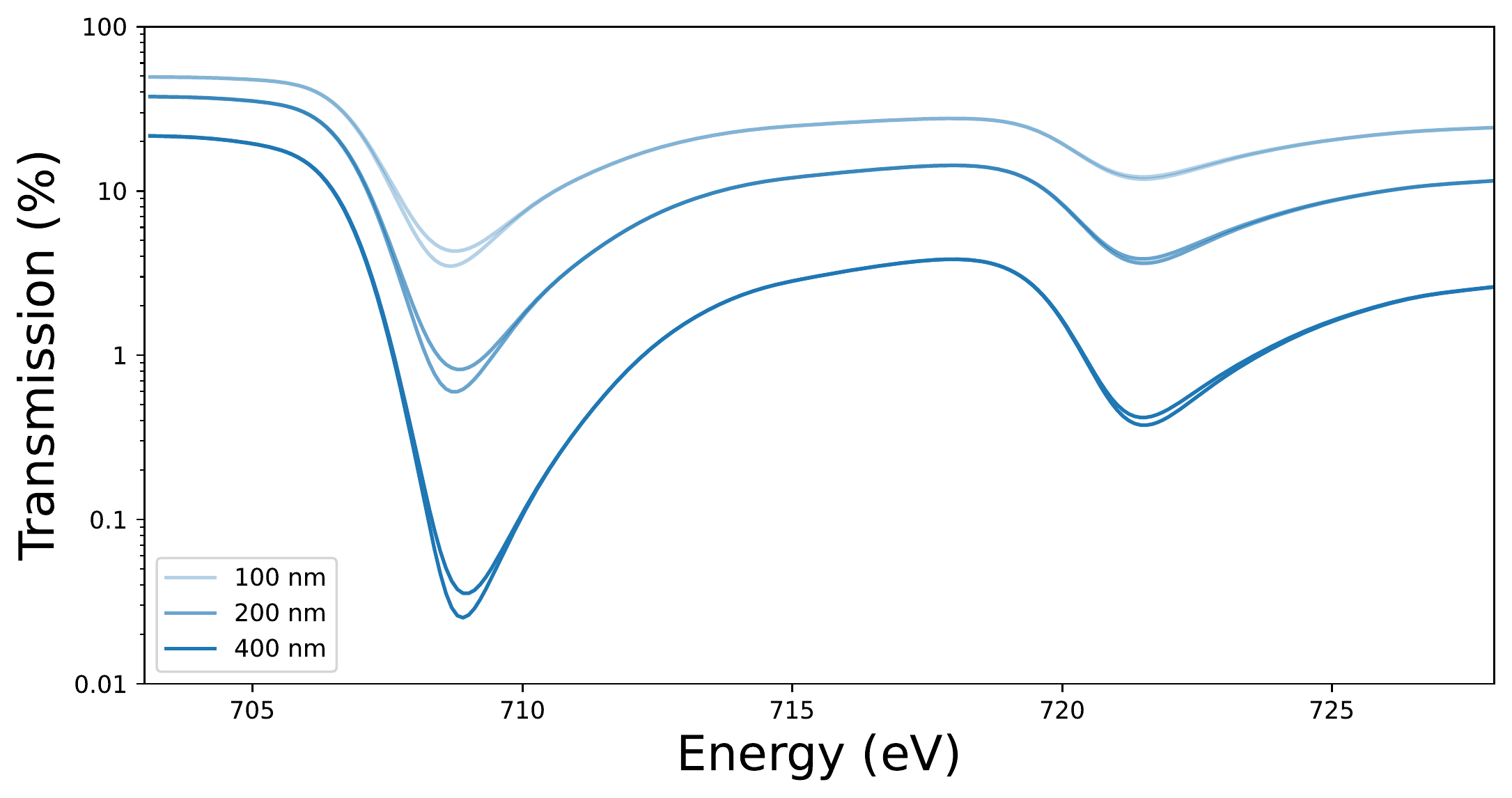}
	\caption{{Calculated pairs of relative transmissions $T^+_\text{tot}(E) = T_\text{tot}(M_\text p, M_\text s=+1, E)$ and $T^-_\text{tot}(E) = T_\text{tot}(M_\text p, M_\text s=-1, E)$, of the initially unpolarized soft x-rays through the polarizer (fixed $M_\text p$) and sample (two sample magnetization states $\pm M_\text s$) for different polarizer thicknesses.
			The simulation has been convoluted my the energy resolution of the setup.}}
	\label{fig:polarizer_total_transmission}
\end{figure}

\begin{figure}[htbp!]
	\centering
	\includegraphics[width=0.75\linewidth]{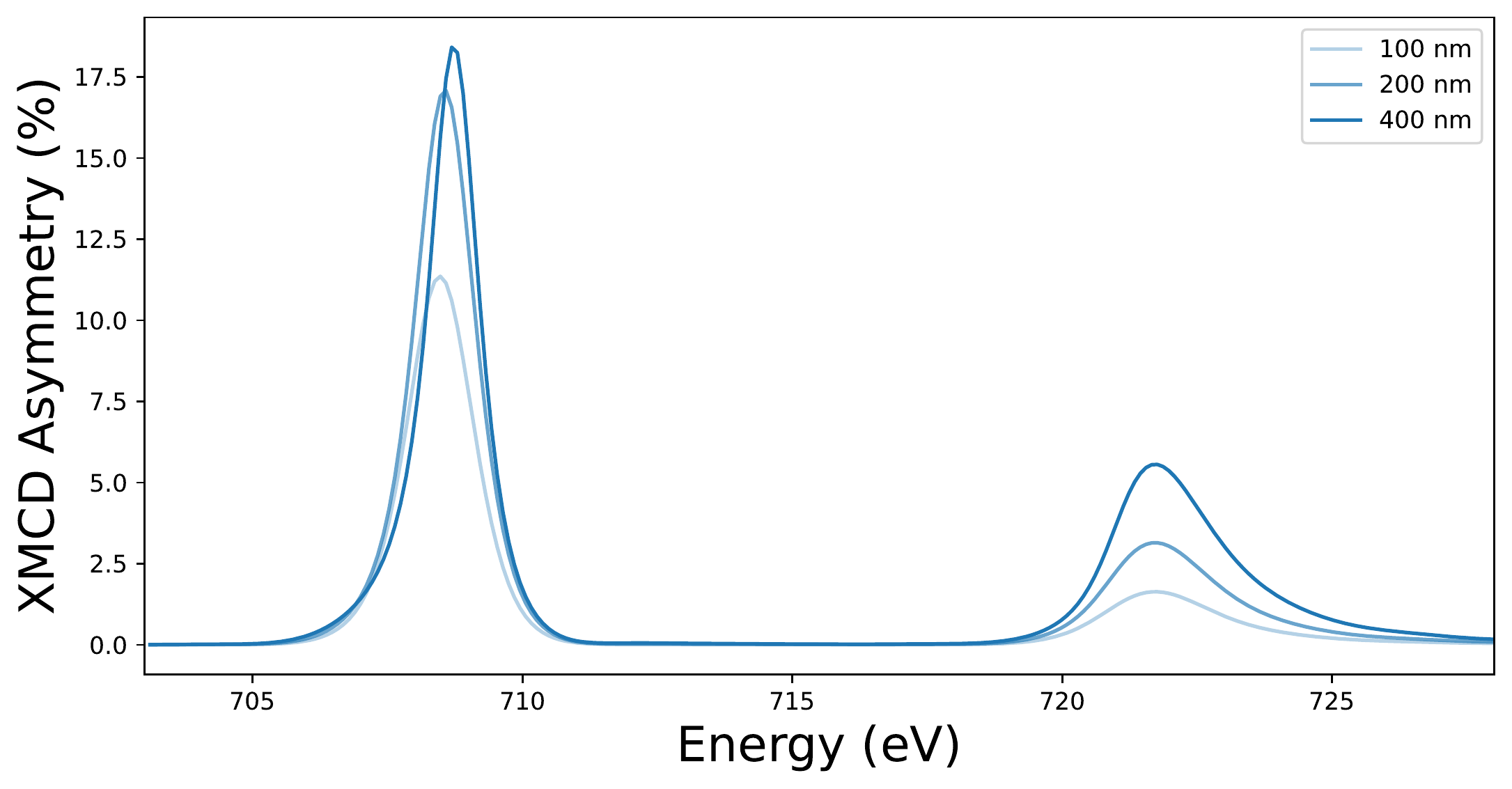}
	\caption{{Simulated asymmetries $A_\text{tot}(M_\text p, M_\text s, E)$ for different thicknesses of the polarizer through the sample.
			The simulation has been convoluted my the energy resolution of the setup.
	}}
	\label{fig:polariser_total_asymmetry}
\end{figure}

For different discrete thicknesses of $d_\text p = [100,200,400]$\,nm and fixed sample thickness $d_\text{s} = 50$\,nm, we simulate the entire dichroic spectrum and the resulting asymmetry across the full energy range.
We convolute this data by the estimated energy resolution of $E/\Delta E \approx550$ of the setup.
The results for the dichroic transmissions are shown in Fig.~\ref{fig:polarizer_total_transmission} and those for the asymmetries in \ref{fig:polariser_total_asymmetry}.

\section{Polarizer Analysis}

\begin{figure}[htbp!]
	\centering
	\includegraphics[width=0.75\linewidth]{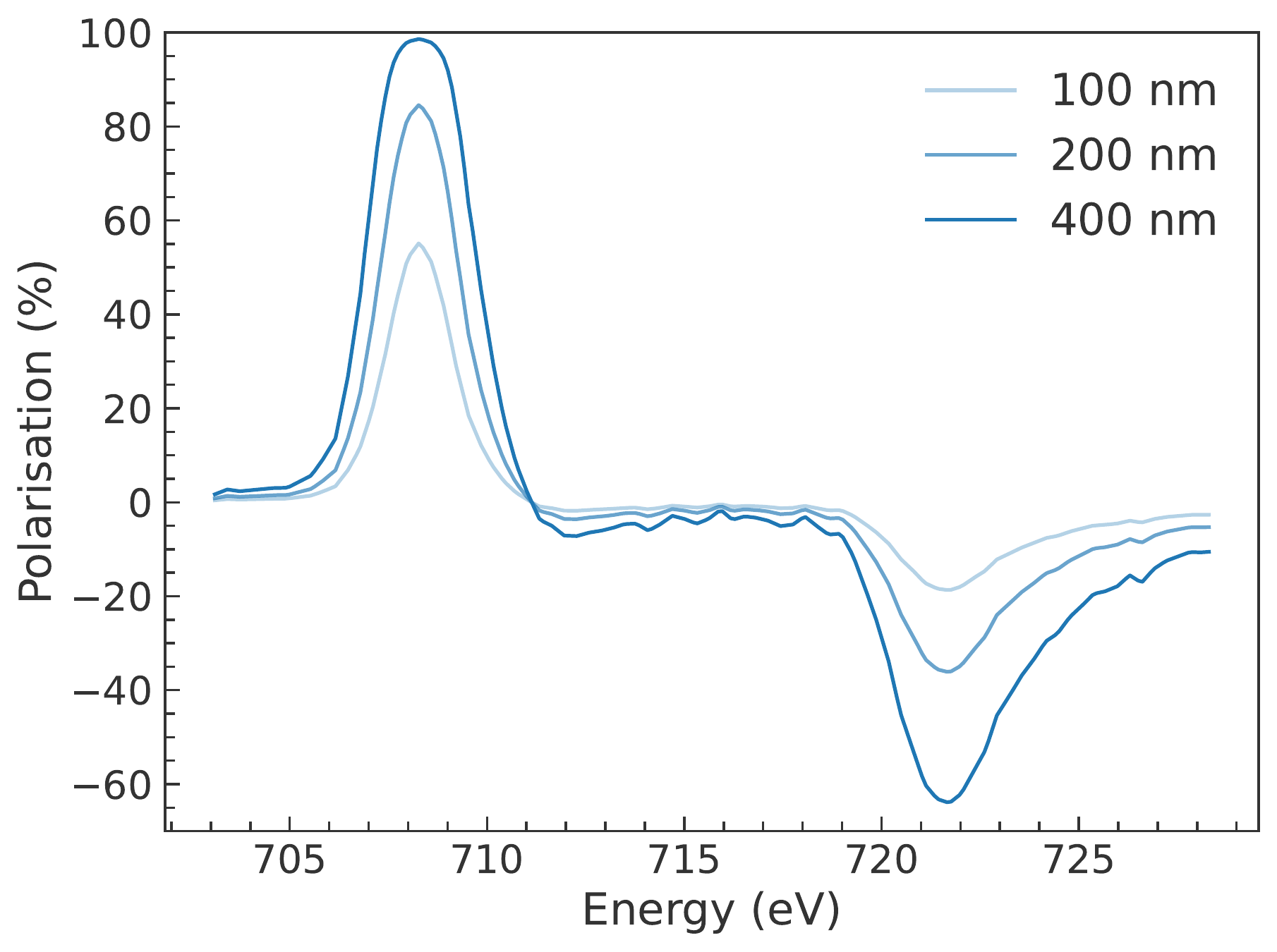}
	\caption{{Degree of polarization for different polarizer thicknesses.
			The spectra are not convoluted with the experimental spectral resolution.}}
	\label{fig:polarization}
\end{figure}

By increasing the thickness of the polarizer, the transmission decreases, but the degree of polarization increases.
In Fig.~\ref{fig:polarization}, we show the expected degree of polarization for a fully magnetized Gd$_{0.24}$Fe$_{0.76}$ film, analogously to the one, discussed in the main text.
Here, we expect a degree of polarization up to 55\% at the Fe $L_3$ edge and -19\% at the Fe $L_2$ edge for the thickness of 100\,nm.
Increasing the thickness to 400\,nm, results in a degree of polarization of 99\% and -64\% at the $L_3$ and $L_2$ edges, respectively.
Note that this calculation is not convoluted by the energy resolution of the RZP.

\section{Synchrotron-radiation beamline measurements}\label{sec:PM3}

\begin{figure}[htbp!]
	\centering
	\includegraphics[width=0.75\linewidth]{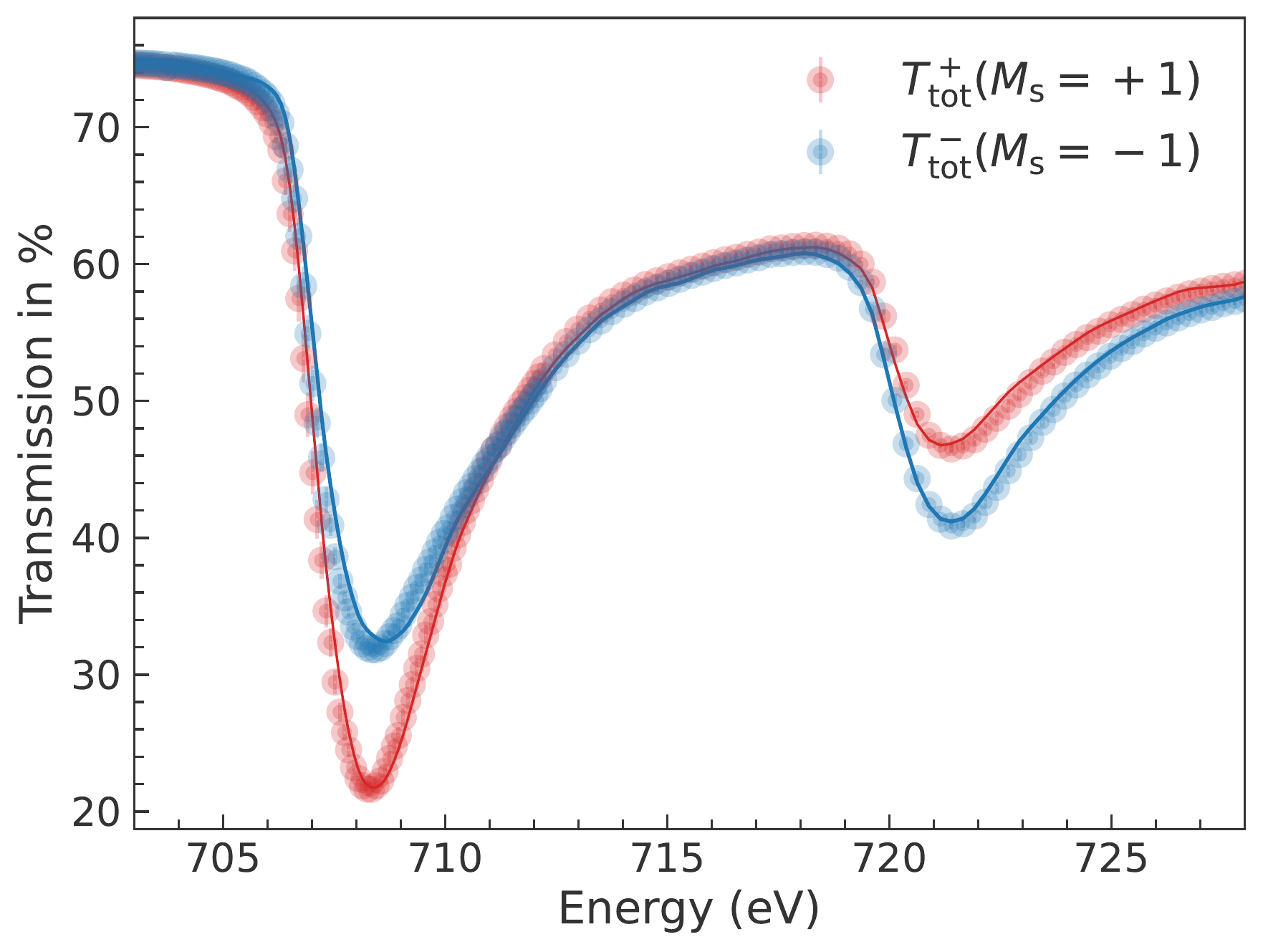}
	\caption{{Dichroic transmissions $T_\text s(M_\text s = +1)$ and $T_\text s(M_\text s = -1)$ through the sample measured at the PM3 beamline of BESSY~II.} Data as dots, simulation as guide to the eye as solid lines.}
	\label{fig:synchrotron_up}
\end{figure}

At the synchrotron-radiation beamline PM3 of BESSY II~\cite{kachel_pm3_2016}, we performed static XMCD transmission experiments on the Gd$_{0.24}$Fe$_{0.76}$ sample to benchmark our simulations.
Here, the polarization state of the soft x-rays can be set to partial circular polarization and the magnetization of the sample can be altered by an electromagnet.
At PM3, the x-ray radiation is monochromatized by the beamline optics and needs to be scanned to acquire XMCD spectra.
For that, we set the energy resolution and step width of the photon energy scans both to approx. 0.25\,eV.
We then integrated for 1\,s at every photon energy with 150 points per full spectrum across the Fe $L$-edges.
This resulted in 150\,s of total integration time per scan per magnetic field direction.
We repeated both scans 6 times which resulted in 30\,min of total integration time for the data as shown in Fig.~\ref{fig:synchrotron_up} together with the best fit results from the magneto-optical simulations.
The large XMCD contrast is a result of the $> 60$\% degree of circular polarization and of the better energy resolution at the bending-magnet beamline.
Note that the laser-based results have been acquired within less than 6\,min of integration time but with 5 times worse energy resolution.
This renders the effective integration times for comparable energy resolution settings quasi equal for both experiments. 
However, the achieved signal-to-noise ratio (SNR) at the plasma x-ray source SNR$_\text{PXS} >$~200 is twice as large as at PM3 with SNR$_\text{PM3} >$~100.

\begin{thebibliography}{46}%
		\makeatletter
		\providecommand \@ifxundefined [1]{%
			\@ifx{#1\undefined}
		}%
		\providecommand \@ifnum [1]{%
			\ifnum #1\expandafter \@firstoftwo
			\else \expandafter \@secondoftwo
			\fi
		}%
		\providecommand \@ifx [1]{%
			\ifx #1\expandafter \@firstoftwo
			\else \expandafter \@secondoftwo
			\fi
		}%
		\providecommand \natexlab [1]{#1}%
		\providecommand \enquote  [1]{``#1''}%
		\providecommand \bibnamefont  [1]{#1}%
		\providecommand \bibfnamefont [1]{#1}%
		\providecommand \citenamefont [1]{#1}%
		\providecommand \href@noop [0]{\@secondoftwo}%
		\providecommand \href [0]{\begingroup \@sanitize@url \@href}%
		\providecommand \@href[1]{\@@startlink{#1}\@@href}%
		\providecommand \@@href[1]{\endgroup#1\@@endlink}%
		\providecommand \@sanitize@url [0]{\catcode `\\12\catcode `\$12\catcode
			`\&12\catcode `\#12\catcode `\^12\catcode `\_12\catcode `\%12\relax}%
		\providecommand \@@startlink[1]{}%
		\providecommand \@@endlink[0]{}%
		\providecommand \url  [0]{\begingroup\@sanitize@url \@url }%
		\providecommand \@url [1]{\endgroup\@href {#1}{\urlprefix }}%
		\providecommand \urlprefix  [0]{URL }%
		\providecommand \Eprint [0]{\href }%
		\providecommand \doibase [0]{https://doi.org/}%
		\providecommand \selectlanguage [0]{\@gobble}%
		\providecommand \bibinfo  [0]{\@secondoftwo}%
		\providecommand \bibfield  [0]{\@secondoftwo}%
		\providecommand \translation [1]{[#1]}%
		\providecommand \BibitemOpen [0]{}%
		\providecommand \bibitemStop [0]{}%
		\providecommand \bibitemNoStop [0]{.\EOS\space}%
		\providecommand \EOS [0]{\spacefactor3000\relax}%
		\providecommand \BibitemShut  [1]{\csname bibitem#1\endcsname}%
		\let\auto@bib@innerbib\@empty
		\bibitem [{\citenamefont {van~der Laan}\ \emph {et~al.}(1986)\citenamefont
			{van~der Laan}, \citenamefont {Thole}, \citenamefont {Sawatzky},
			\citenamefont {Goedkoop}, \citenamefont {Fuggle}, \citenamefont {Esteva},
			\citenamefont {Karnatak}, \citenamefont {Remeika},\ and\ \citenamefont
			{Dabkowska}}]{VanderLaan1986}%
		\BibitemOpen
		\bibfield  {author} {\bibinfo {author} {\bibfnamefont {G.}~\bibnamefont
				{van~der Laan}}, \bibinfo {author} {\bibfnamefont {B.~T.}\ \bibnamefont
				{Thole}}, \bibinfo {author} {\bibfnamefont {G.~A.}\ \bibnamefont {Sawatzky}},
			\bibinfo {author} {\bibfnamefont {J.~B.}\ \bibnamefont {Goedkoop}}, \bibinfo
			{author} {\bibfnamefont {J.~C.}\ \bibnamefont {Fuggle}}, \bibinfo {author}
			{\bibfnamefont {J.-M.}\ \bibnamefont {Esteva}}, \bibinfo {author}
			{\bibfnamefont {R.}~\bibnamefont {Karnatak}}, \bibinfo {author}
			{\bibfnamefont {J.~P.}\ \bibnamefont {Remeika}},\ and\ \bibinfo {author}
			{\bibfnamefont {H.~A.}\ \bibnamefont {Dabkowska}},\ }\bibfield  {title}
		{\bibinfo {title} {{Experimental proof of magnetic x-ray dichroism}},\ }\href
		{https://doi.org/10.1103/PhysRevB.34.6529} {\bibfield  {journal} {\bibinfo
				{journal} {Phys. Rev. B}\ }\textbf {\bibinfo {volume} {34}},\ \bibinfo
			{pages} {6529} (\bibinfo {year} {1986})}\BibitemShut {NoStop}%
		\bibitem [{\citenamefont {Sch{\"{u}}tz}\ \emph {et~al.}(1987)\citenamefont
			{Sch{\"{u}}tz}, \citenamefont {Wagner}, \citenamefont {Wilhelm},
			\citenamefont {Kienle}, \citenamefont {Zeller}, \citenamefont {Frahm},\ and\
			\citenamefont {Materlik}}]{Schutz1987}%
		\BibitemOpen
		\bibfield  {author} {\bibinfo {author} {\bibfnamefont {G.}~\bibnamefont
				{Sch{\"{u}}tz}}, \bibinfo {author} {\bibfnamefont {W.}~\bibnamefont
				{Wagner}}, \bibinfo {author} {\bibfnamefont {W.}~\bibnamefont {Wilhelm}},
			\bibinfo {author} {\bibfnamefont {P.}~\bibnamefont {Kienle}}, \bibinfo
			{author} {\bibfnamefont {R.}~\bibnamefont {Zeller}}, \bibinfo {author}
			{\bibfnamefont {R.}~\bibnamefont {Frahm}},\ and\ \bibinfo {author}
			{\bibfnamefont {G.}~\bibnamefont {Materlik}},\ }\bibfield  {title} {\bibinfo
			{title} {{Absorption of circularly polarized x rays in iron}},\ }\href
		{https://doi.org/10.1103/PhysRevLett.58.737} {\bibfield  {journal} {\bibinfo
				{journal} {Phys. Rev. Lett.}\ }\textbf {\bibinfo {volume} {58}},\ \bibinfo
			{pages} {737} (\bibinfo {year} {1987})}\BibitemShut {NoStop}%
		\bibitem [{\citenamefont {St{\"{o}}hr}(1999)}]{Stohr1999}%
		\BibitemOpen
		\bibfield  {author} {\bibinfo {author} {\bibfnamefont {J.}~\bibnamefont
				{St{\"{o}}hr}},\ }\bibfield  {title} {\bibinfo {title} {{Exploring the
					microscopic origin of magnetic anisotropies with X-ray magnetic circular
					dichroism (XMCD) spectroscopy}},\ }\href
		{https://doi.org/10.1016/S0304-8853(99)00407-2} {\bibfield  {journal}
			{\bibinfo  {journal} {J. Magn. Magn. Mater.}\ }\textbf {\bibinfo {volume}
				{200}},\ \bibinfo {pages} {470} (\bibinfo {year} {1999})}\BibitemShut
		{NoStop}%
		\bibitem [{\citenamefont {van~der Laan}\ and\ \citenamefont
			{Figueroa}(2014)}]{VanderLaan2014}%
		\BibitemOpen
		\bibfield  {author} {\bibinfo {author} {\bibfnamefont {G.}~\bibnamefont
				{van~der Laan}}\ and\ \bibinfo {author} {\bibfnamefont {A.~I.}\ \bibnamefont
				{Figueroa}},\ }\bibfield  {title} {\bibinfo {title} {{X-ray magnetic circular
					dichroism—A versatile tool to study magnetism}},\ }\href
		{https://doi.org/10.1016/j.ccr.2014.03.018} {\bibfield  {journal} {\bibinfo
				{journal} {Coord. Chem. Rev.}\ }\textbf {\bibinfo {volume} {277-278}},\
			\bibinfo {pages} {95} (\bibinfo {year} {2014})}\BibitemShut {NoStop}%
		\bibitem [{\citenamefont {Kortright}\ \emph {et~al.}(1999)\citenamefont
			{Kortright}, \citenamefont {Awschalom}, \citenamefont {St{\"{o}}hr},
			\citenamefont {Bader}, \citenamefont {Idzerda}, \citenamefont {Parkin},
			\citenamefont {Schuller},\ and\ \citenamefont {Siegmann}}]{Kortright1999}%
		\BibitemOpen
		\bibfield  {author} {\bibinfo {author} {\bibfnamefont {J.~B.}\ \bibnamefont
				{Kortright}}, \bibinfo {author} {\bibfnamefont {D.~D.}\ \bibnamefont
				{Awschalom}}, \bibinfo {author} {\bibfnamefont {J.}~\bibnamefont
				{St{\"{o}}hr}}, \bibinfo {author} {\bibfnamefont {S.~D.}\ \bibnamefont
				{Bader}}, \bibinfo {author} {\bibfnamefont {Y.~U.}\ \bibnamefont {Idzerda}},
			\bibinfo {author} {\bibfnamefont {S.~S.}\ \bibnamefont {Parkin}}, \bibinfo
			{author} {\bibfnamefont {I.~K.}\ \bibnamefont {Schuller}},\ and\ \bibinfo
			{author} {\bibfnamefont {H.~C.}\ \bibnamefont {Siegmann}},\ }\bibfield
		{title} {\bibinfo {title} {{Research frontiers in magnetic materials at soft
					X-ray synchrotron radiation facilities}},\ }\href
		{https://doi.org/10.1016/S0304-8853(99)00485-0} {\bibfield  {journal}
			{\bibinfo  {journal} {J. Magn. Magn. Mater.}\ }\textbf {\bibinfo {volume}
				{207}},\ \bibinfo {pages} {7} (\bibinfo {year} {1999})}\BibitemShut {NoStop}%
		\bibitem [{\citenamefont {Thole}\ \emph {et~al.}(1992)\citenamefont {Thole},
			\citenamefont {Carra}, \citenamefont {Sette},\ and\ \citenamefont {van~der
				Laan}}]{Thole1992}%
		\BibitemOpen
		\bibfield  {author} {\bibinfo {author} {\bibfnamefont {B.~T.}\ \bibnamefont
				{Thole}}, \bibinfo {author} {\bibfnamefont {P.}~\bibnamefont {Carra}},
			\bibinfo {author} {\bibfnamefont {F.}~\bibnamefont {Sette}},\ and\ \bibinfo
			{author} {\bibfnamefont {G.}~\bibnamefont {van~der Laan}},\ }\bibfield
		{title} {\bibinfo {title} {{X-ray circular dichroism as a probe of orbital
					magnetization}},\ }\href {https://doi.org/10.1103/PhysRevLett.68.1943}
		{\bibfield  {journal} {\bibinfo  {journal} {Phys. Rev. Lett.}\ }\textbf
			{\bibinfo {volume} {68}},\ \bibinfo {pages} {1943} (\bibinfo {year}
			{1992})}\BibitemShut {NoStop}%
		\bibitem [{\citenamefont {Carra}\ \emph {et~al.}(1993)\citenamefont {Carra},
			\citenamefont {Thole}, \citenamefont {Altarelli},\ and\ \citenamefont
			{Wang}}]{Carra1993}%
		\BibitemOpen
		\bibfield  {author} {\bibinfo {author} {\bibfnamefont {P.}~\bibnamefont
				{Carra}}, \bibinfo {author} {\bibfnamefont {B.~T.}\ \bibnamefont {Thole}},
			\bibinfo {author} {\bibfnamefont {M.}~\bibnamefont {Altarelli}},\ and\
			\bibinfo {author} {\bibfnamefont {X.}~\bibnamefont {Wang}},\ }\bibfield
		{title} {\bibinfo {title} {{X-ray circular dichroism and local magnetic
					fields}},\ }\href {https://doi.org/10.1103/PhysRevLett.70.694} {\bibfield
			{journal} {\bibinfo  {journal} {Phys. Rev. Lett.}\ }\textbf {\bibinfo
				{volume} {70}},\ \bibinfo {pages} {694} (\bibinfo {year} {1993})}\BibitemShut
		{NoStop}%
		\bibitem [{\citenamefont {St{\"{o}}hr}\ and\ \citenamefont
			{K{\"{o}}nig}(1995)}]{Stohr1995}%
		\BibitemOpen
		\bibfield  {author} {\bibinfo {author} {\bibfnamefont {J.}~\bibnamefont
				{St{\"{o}}hr}}\ and\ \bibinfo {author} {\bibfnamefont {H.}~\bibnamefont
				{K{\"{o}}nig}},\ }\bibfield  {title} {\bibinfo {title} {{Determination of
					Spin- and Orbital-Moment Anisotropies in Transition Metals by Angle-Dependent
					X-Ray Magnetic Circular Dichroism}},\ }\href
		{https://doi.org/10.1103/PhysRevLett.75.3748} {\bibfield  {journal} {\bibinfo
				{journal} {Phys. Rev. Lett.}\ }\textbf {\bibinfo {volume} {75}},\ \bibinfo
			{pages} {3748} (\bibinfo {year} {1995})}\BibitemShut {NoStop}%
		\bibitem [{\citenamefont {St{\"{o}}hr}\ \emph {et~al.}(1993)\citenamefont
			{St{\"{o}}hr}, \citenamefont {Wu}, \citenamefont {Hermsmeier}, \citenamefont
			{Samant}, \citenamefont {Harp}, \citenamefont {Koranda}, \citenamefont
			{Dunham},\ and\ \citenamefont {Tonner}}]{Stohr1993}%
		\BibitemOpen
		\bibfield  {author} {\bibinfo {author} {\bibfnamefont {J.}~\bibnamefont
				{St{\"{o}}hr}}, \bibinfo {author} {\bibfnamefont {Y.}~\bibnamefont {Wu}},
			\bibinfo {author} {\bibfnamefont {B.~D.}\ \bibnamefont {Hermsmeier}},
			\bibinfo {author} {\bibfnamefont {M.~G.}\ \bibnamefont {Samant}}, \bibinfo
			{author} {\bibfnamefont {G.~R.}\ \bibnamefont {Harp}}, \bibinfo {author}
			{\bibfnamefont {S.}~\bibnamefont {Koranda}}, \bibinfo {author} {\bibfnamefont
				{D.}~\bibnamefont {Dunham}},\ and\ \bibinfo {author} {\bibfnamefont {B.~P.}\
				\bibnamefont {Tonner}},\ }\bibfield  {title} {\bibinfo {title}
			{{Element-Specific Magnetic Microscopy with Circularly Polarized X-rays}},\
		}\href {https://doi.org/10.1126/science.259.5095.658} {\bibfield  {journal}
			{\bibinfo  {journal} {Science (80-. ).}\ }\textbf {\bibinfo {volume} {259}},\
			\bibinfo {pages} {658} (\bibinfo {year} {1993})}\BibitemShut {NoStop}%
		\bibitem [{\citenamefont {Eisebitt}\ \emph {et~al.}(2004)\citenamefont
			{Eisebitt}, \citenamefont {L{\"{u}}ning}, \citenamefont {Schlotter},
			\citenamefont {L{\"{o}}rgen}, \citenamefont {Hellwig}, \citenamefont
			{Eberhardt},\ and\ \citenamefont {St{\"{o}}hr}}]{Eisebitt2004}%
		\BibitemOpen
		\bibfield  {author} {\bibinfo {author} {\bibfnamefont {S.}~\bibnamefont
				{Eisebitt}}, \bibinfo {author} {\bibfnamefont {J.}~\bibnamefont
				{L{\"{u}}ning}}, \bibinfo {author} {\bibfnamefont {W.~F.}\ \bibnamefont
				{Schlotter}}, \bibinfo {author} {\bibfnamefont {M.}~\bibnamefont
				{L{\"{o}}rgen}}, \bibinfo {author} {\bibfnamefont {O.}~\bibnamefont
				{Hellwig}}, \bibinfo {author} {\bibfnamefont {W.}~\bibnamefont {Eberhardt}},\
			and\ \bibinfo {author} {\bibfnamefont {J.}~\bibnamefont {St{\"{o}}hr}},\
		}\bibfield  {title} {\bibinfo {title} {{Lensless imaging of magnetic
					nanostructures by X-ray spectro-holography}},\ }\href
		{https://doi.org/10.1038/nature03139} {\bibfield  {journal} {\bibinfo
				{journal} {Nature}\ }\textbf {\bibinfo {volume} {432}},\ \bibinfo {pages}
			{885} (\bibinfo {year} {2004})}\BibitemShut {NoStop}%
		\bibitem [{\citenamefont {Higley}\ \emph {et~al.}(2019)\citenamefont {Higley},
			\citenamefont {Reid}, \citenamefont {Chen}, \citenamefont {Guyader},
			\citenamefont {Hellwig}, \citenamefont {Lutman}, \citenamefont {Liu},
			\citenamefont {Shafer}, \citenamefont {Chase}, \citenamefont {Dakovski},
			\citenamefont {Mitra}, \citenamefont {Yuan}, \citenamefont {Schlappa},
			\citenamefont {D{\"{u}}rr}, \citenamefont {Schlotter},\ and\ \citenamefont
			{St{\"{o}}hr}}]{Higley2019}%
		\BibitemOpen
		\bibfield  {author} {\bibinfo {author} {\bibfnamefont {D.~J.}\ \bibnamefont
				{Higley}}, \bibinfo {author} {\bibfnamefont {A.~H.}\ \bibnamefont {Reid}},
			\bibinfo {author} {\bibfnamefont {Z.}~\bibnamefont {Chen}}, \bibinfo {author}
			{\bibfnamefont {L.~L.}\ \bibnamefont {Guyader}}, \bibinfo {author}
			{\bibfnamefont {O.}~\bibnamefont {Hellwig}}, \bibinfo {author} {\bibfnamefont
				{A.~A.}\ \bibnamefont {Lutman}}, \bibinfo {author} {\bibfnamefont
				{T.}~\bibnamefont {Liu}}, \bibinfo {author} {\bibfnamefont {P.}~\bibnamefont
				{Shafer}}, \bibinfo {author} {\bibfnamefont {T.}~\bibnamefont {Chase}},
			\bibinfo {author} {\bibfnamefont {G.~L.}\ \bibnamefont {Dakovski}}, \bibinfo
			{author} {\bibfnamefont {A.}~\bibnamefont {Mitra}}, \bibinfo {author}
			{\bibfnamefont {E.}~\bibnamefont {Yuan}}, \bibinfo {author} {\bibfnamefont
				{J.}~\bibnamefont {Schlappa}}, \bibinfo {author} {\bibfnamefont {H.~A.}\
				\bibnamefont {D{\"{u}}rr}}, \bibinfo {author} {\bibfnamefont {W.~F.}\
				\bibnamefont {Schlotter}},\ and\ \bibinfo {author} {\bibfnamefont
				{J.}~\bibnamefont {St{\"{o}}hr}},\ }\bibfield  {title} {\bibinfo {title}
			{{Femtosecond X-ray induced changes of the electronic and magnetic response
					of solids from electron redistribution}},\ }\href
		{https://doi.org/10.1038/s41467-019-13272-5} {\bibfield  {journal} {\bibinfo
				{journal} {Nat. Commun.}\ }\textbf {\bibinfo {volume} {10}},\ \bibinfo
			{pages} {5289} (\bibinfo {year} {2019})}\BibitemShut {NoStop}%
		\bibitem [{\citenamefont {Holldack}\ \emph {et~al.}(2014)\citenamefont
			{Holldack}, \citenamefont {Bahrdt}, \citenamefont {Balzer}, \citenamefont
			{Bovensiepen}, \citenamefont {Brzhezinskaya}, \citenamefont {Erko},
			\citenamefont {Eschenlohr}, \citenamefont {Follath}, \citenamefont {Firsov},
			\citenamefont {Frentrup}, \citenamefont {{Le Guyader}}, \citenamefont
			{Kachel}, \citenamefont {Kuske}, \citenamefont {Mitzner}, \citenamefont
			{M{\"{u}}ller}, \citenamefont {Pontius}, \citenamefont {Quast}, \citenamefont
			{Radu}, \citenamefont {Schmidt}, \citenamefont
			{Sch{\"{u}}{\ss}ler-Langeheine}, \citenamefont {Sperling}, \citenamefont
			{Stamm}, \citenamefont {Trabant},\ and\ \citenamefont
			{F{\"{o}}hlisch}}]{Holldack2014}%
		\BibitemOpen
		\bibfield  {author} {\bibinfo {author} {\bibfnamefont {K.}~\bibnamefont
				{Holldack}}, \bibinfo {author} {\bibfnamefont {J.}~\bibnamefont {Bahrdt}},
			\bibinfo {author} {\bibfnamefont {A.}~\bibnamefont {Balzer}}, \bibinfo
			{author} {\bibfnamefont {U.}~\bibnamefont {Bovensiepen}}, \bibinfo {author}
			{\bibfnamefont {M.}~\bibnamefont {Brzhezinskaya}}, \bibinfo {author}
			{\bibfnamefont {A.}~\bibnamefont {Erko}}, \bibinfo {author} {\bibfnamefont
				{A.}~\bibnamefont {Eschenlohr}}, \bibinfo {author} {\bibfnamefont
				{R.}~\bibnamefont {Follath}}, \bibinfo {author} {\bibfnamefont
				{A.}~\bibnamefont {Firsov}}, \bibinfo {author} {\bibfnamefont
				{W.}~\bibnamefont {Frentrup}}, \bibinfo {author} {\bibfnamefont
				{L.}~\bibnamefont {{Le Guyader}}}, \bibinfo {author} {\bibfnamefont
				{T.}~\bibnamefont {Kachel}}, \bibinfo {author} {\bibfnamefont
				{P.}~\bibnamefont {Kuske}}, \bibinfo {author} {\bibfnamefont
				{R.}~\bibnamefont {Mitzner}}, \bibinfo {author} {\bibfnamefont
				{R.}~\bibnamefont {M{\"{u}}ller}}, \bibinfo {author} {\bibfnamefont
				{N.}~\bibnamefont {Pontius}}, \bibinfo {author} {\bibfnamefont
				{T.}~\bibnamefont {Quast}}, \bibinfo {author} {\bibfnamefont
				{I.}~\bibnamefont {Radu}}, \bibinfo {author} {\bibfnamefont {J.-S.}\
				\bibnamefont {Schmidt}}, \bibinfo {author} {\bibfnamefont {C.}~\bibnamefont
				{Sch{\"{u}}{\ss}ler-Langeheine}}, \bibinfo {author} {\bibfnamefont
				{M.}~\bibnamefont {Sperling}}, \bibinfo {author} {\bibfnamefont
				{C.}~\bibnamefont {Stamm}}, \bibinfo {author} {\bibfnamefont
				{C.}~\bibnamefont {Trabant}},\ and\ \bibinfo {author} {\bibfnamefont
				{A.}~\bibnamefont {F{\"{o}}hlisch}},\ }\bibfield  {title} {\bibinfo {title}
			{{FemtoSpeX: a versatile optical pump–soft X-ray probe facility with 100 fs
					X-ray pulses of variable polarization}},\ }\href
		{https://doi.org/10.1107/S1600577514012247} {\bibfield  {journal} {\bibinfo
				{journal} {J. Synchrotron Radiat.}\ }\textbf {\bibinfo {volume} {21}},\
			\bibinfo {pages} {1090} (\bibinfo {year} {2014})}\BibitemShut {NoStop}%
		\bibitem [{\citenamefont {Siegrist}\ \emph {et~al.}(2019)\citenamefont
			{Siegrist}, \citenamefont {Gessner}, \citenamefont {Ossiander}, \citenamefont
			{Denker}, \citenamefont {Chang}, \citenamefont {Schr{\"{o}}der},
			\citenamefont {Guggenmos}, \citenamefont {Cui}, \citenamefont {Walowski},
			\citenamefont {Martens}, \citenamefont {Dewhurst}, \citenamefont
			{Kleineberg}, \citenamefont {M{\"{u}}nzenberg}, \citenamefont {Sharma},\ and\
			\citenamefont {Schultze}}]{Siegrist2019}%
		\BibitemOpen
		\bibfield  {author} {\bibinfo {author} {\bibfnamefont {F.}~\bibnamefont
				{Siegrist}}, \bibinfo {author} {\bibfnamefont {J.~A.}\ \bibnamefont
				{Gessner}}, \bibinfo {author} {\bibfnamefont {M.}~\bibnamefont {Ossiander}},
			\bibinfo {author} {\bibfnamefont {C.}~\bibnamefont {Denker}}, \bibinfo
			{author} {\bibfnamefont {Y.-P.}\ \bibnamefont {Chang}}, \bibinfo {author}
			{\bibfnamefont {M.~C.}\ \bibnamefont {Schr{\"{o}}der}}, \bibinfo {author}
			{\bibfnamefont {A.}~\bibnamefont {Guggenmos}}, \bibinfo {author}
			{\bibfnamefont {Y.}~\bibnamefont {Cui}}, \bibinfo {author} {\bibfnamefont
				{J.}~\bibnamefont {Walowski}}, \bibinfo {author} {\bibfnamefont
				{U.}~\bibnamefont {Martens}}, \bibinfo {author} {\bibfnamefont {J.~K.}\
				\bibnamefont {Dewhurst}}, \bibinfo {author} {\bibfnamefont {U.}~\bibnamefont
				{Kleineberg}}, \bibinfo {author} {\bibfnamefont {M.}~\bibnamefont
				{M{\"{u}}nzenberg}}, \bibinfo {author} {\bibfnamefont {S.}~\bibnamefont
				{Sharma}},\ and\ \bibinfo {author} {\bibfnamefont {M.}~\bibnamefont
				{Schultze}},\ }\bibfield  {title} {\bibinfo {title} {{Light-wave dynamic
					control of magnetism}},\ }\href {https://doi.org/10.1038/s41586-019-1333-x}
		{\bibfield  {journal} {\bibinfo  {journal} {Nature}\ }\textbf {\bibinfo
				{volume} {571}},\ \bibinfo {pages} {240} (\bibinfo {year}
			{2019})}\BibitemShut {NoStop}%
		\bibitem [{\citenamefont {Willems}\ \emph {et~al.}(2020)\citenamefont
			{Willems}, \citenamefont {{von Korff Schmising}}, \citenamefont
			{Str{\"{u}}ber}, \citenamefont {Schick}, \citenamefont {Engel}, \citenamefont
			{Dewhurst}, \citenamefont {Elliott}, \citenamefont {Sharma},\ and\
			\citenamefont {Eisebitt}}]{Willems2020}%
		\BibitemOpen
		\bibfield  {author} {\bibinfo {author} {\bibfnamefont {F.}~\bibnamefont
				{Willems}}, \bibinfo {author} {\bibfnamefont {C.}~\bibnamefont {{von Korff
						Schmising}}}, \bibinfo {author} {\bibfnamefont {C.}~\bibnamefont
				{Str{\"{u}}ber}}, \bibinfo {author} {\bibfnamefont {D.}~\bibnamefont
				{Schick}}, \bibinfo {author} {\bibfnamefont {D.~W.}\ \bibnamefont {Engel}},
			\bibinfo {author} {\bibfnamefont {J.~K.}\ \bibnamefont {Dewhurst}}, \bibinfo
			{author} {\bibfnamefont {P.}~\bibnamefont {Elliott}}, \bibinfo {author}
			{\bibfnamefont {S.}~\bibnamefont {Sharma}},\ and\ \bibinfo {author}
			{\bibfnamefont {S.}~\bibnamefont {Eisebitt}},\ }\bibfield  {title} {\bibinfo
			{title} {{Optical inter-site spin transfer probed by energy and spin-resolved
					transient absorption spectroscopy}},\ }\href
		{https://doi.org/10.1038/s41467-020-14691-5} {\bibfield  {journal} {\bibinfo
				{journal} {Nat. Commun.}\ }\textbf {\bibinfo {volume} {11}},\ \bibinfo
			{pages} {871} (\bibinfo {year} {2020})}\BibitemShut {NoStop}%
		\bibitem [{\citenamefont {Hofherr}\ \emph {et~al.}(2020)\citenamefont
			{Hofherr}, \citenamefont {Häuser}, \citenamefont {Dewhurst}, \citenamefont
			{Tengdin}, \citenamefont {Sakshath}, \citenamefont {Nembach}, \citenamefont
			{Weber}, \citenamefont {Shaw}, \citenamefont {Silva}, \citenamefont
			{Kapteyn}, \citenamefont {Cinchetti}, \citenamefont {Rethfeld}, \citenamefont
			{Murnane}, \citenamefont {Steil}, \citenamefont {Stadtmüller}, \citenamefont
			{Sharma}, \citenamefont {Aeschlimann},\ and\ \citenamefont
			{Mathias}}]{hofherr_ultrafast_2020}%
		\BibitemOpen
		\bibfield  {author} {\bibinfo {author} {\bibfnamefont {M.}~\bibnamefont
				{Hofherr}}, \bibinfo {author} {\bibfnamefont {S.}~\bibnamefont {Häuser}},
			\bibinfo {author} {\bibfnamefont {J.~K.}\ \bibnamefont {Dewhurst}}, \bibinfo
			{author} {\bibfnamefont {P.}~\bibnamefont {Tengdin}}, \bibinfo {author}
			{\bibfnamefont {S.}~\bibnamefont {Sakshath}}, \bibinfo {author}
			{\bibfnamefont {H.~T.}\ \bibnamefont {Nembach}}, \bibinfo {author}
			{\bibfnamefont {S.~T.}\ \bibnamefont {Weber}}, \bibinfo {author}
			{\bibfnamefont {J.~M.}\ \bibnamefont {Shaw}}, \bibinfo {author}
			{\bibfnamefont {T.~J.}\ \bibnamefont {Silva}}, \bibinfo {author}
			{\bibfnamefont {H.~C.}\ \bibnamefont {Kapteyn}}, \bibinfo {author}
			{\bibfnamefont {M.}~\bibnamefont {Cinchetti}}, \bibinfo {author}
			{\bibfnamefont {B.}~\bibnamefont {Rethfeld}}, \bibinfo {author}
			{\bibfnamefont {M.~M.}\ \bibnamefont {Murnane}}, \bibinfo {author}
			{\bibfnamefont {D.}~\bibnamefont {Steil}}, \bibinfo {author} {\bibfnamefont
				{B.}~\bibnamefont {Stadtmüller}}, \bibinfo {author} {\bibfnamefont
				{S.}~\bibnamefont {Sharma}}, \bibinfo {author} {\bibfnamefont
				{M.}~\bibnamefont {Aeschlimann}},\ and\ \bibinfo {author} {\bibfnamefont
				{S.}~\bibnamefont {Mathias}},\ }\bibfield  {title} {\bibinfo {title}
			{Ultrafast optically induced spin transfer in ferromagnetic alloys},\ }\href
		{https://doi.org/10.1126/sciadv.aay8717} {\bibfield  {journal} {\bibinfo
				{journal} {Sci. Adv.}\ }\textbf {\bibinfo {volume} {6}},\ \bibinfo {pages}
			{eaay8717} (\bibinfo {year} {2020})}\BibitemShut {NoStop}%
		\bibitem [{\citenamefont {Kfir}\ \emph {et~al.}(2015)\citenamefont {Kfir},
			\citenamefont {Grychtol}, \citenamefont {Turgut}, \citenamefont {Knut},
			\citenamefont {Zusin}, \citenamefont {Popmintchev}, \citenamefont
			{Popmintchev}, \citenamefont {Nembach}, \citenamefont {Shaw}, \citenamefont
			{Fleischer}, \citenamefont {Kapteyn}, \citenamefont {Murnane},\ and\
			\citenamefont {Cohen}}]{Kfir2015}%
		\BibitemOpen
		\bibfield  {author} {\bibinfo {author} {\bibfnamefont {O.}~\bibnamefont
				{Kfir}}, \bibinfo {author} {\bibfnamefont {P.}~\bibnamefont {Grychtol}},
			\bibinfo {author} {\bibfnamefont {E.}~\bibnamefont {Turgut}}, \bibinfo
			{author} {\bibfnamefont {R.}~\bibnamefont {Knut}}, \bibinfo {author}
			{\bibfnamefont {D.}~\bibnamefont {Zusin}}, \bibinfo {author} {\bibfnamefont
				{D.}~\bibnamefont {Popmintchev}}, \bibinfo {author} {\bibfnamefont
				{T.}~\bibnamefont {Popmintchev}}, \bibinfo {author} {\bibfnamefont
				{H.}~\bibnamefont {Nembach}}, \bibinfo {author} {\bibfnamefont {J.~M.}\
				\bibnamefont {Shaw}}, \bibinfo {author} {\bibfnamefont {A.}~\bibnamefont
				{Fleischer}}, \bibinfo {author} {\bibfnamefont {H.}~\bibnamefont {Kapteyn}},
			\bibinfo {author} {\bibfnamefont {M.}~\bibnamefont {Murnane}},\ and\ \bibinfo
			{author} {\bibfnamefont {O.}~\bibnamefont {Cohen}},\ }\bibfield  {title}
		{\bibinfo {title} {{Generation of bright phase-matched circularly-polarized
					extreme ultraviolet high harmonics}},\ }\href
		{https://doi.org/10.1038/nphoton.2014.293} {\bibfield  {journal} {\bibinfo
				{journal} {Nat. Photonics}\ }\textbf {\bibinfo {volume} {9}},\ \bibinfo
			{pages} {99} (\bibinfo {year} {2015})}\BibitemShut {NoStop}%
		\bibitem [{\citenamefont {Fan}\ \emph {et~al.}(2015)\citenamefont {Fan},
			\citenamefont {Grychtol}, \citenamefont {Knut}, \citenamefont
			{Hern{\'{a}}ndez-Garc{\'{i}}a}, \citenamefont {Hickstein}, \citenamefont
			{Zusin}, \citenamefont {Gentry}, \citenamefont {Dollar}, \citenamefont
			{Mancuso}, \citenamefont {Hogle}, \citenamefont {Kfir}, \citenamefont
			{Legut}, \citenamefont {Carva}, \citenamefont {Ellis}, \citenamefont
			{Dorney}, \citenamefont {Chen}, \citenamefont {Shpyrko}, \citenamefont
			{Fullerton}, \citenamefont {Cohen}, \citenamefont {Oppeneer}, \citenamefont
			{Milo{\v{s}}evi{\'{c}}}, \citenamefont {Becker}, \citenamefont
			{Jaro{\'{n}}-Becker}, \citenamefont {Popmintchev}, \citenamefont {Murnane},\
			and\ \citenamefont {Kapteyn}}]{Fan2015}%
		\BibitemOpen
		\bibfield  {author} {\bibinfo {author} {\bibfnamefont {T.}~\bibnamefont
				{Fan}}, \bibinfo {author} {\bibfnamefont {P.}~\bibnamefont {Grychtol}},
			\bibinfo {author} {\bibfnamefont {R.}~\bibnamefont {Knut}}, \bibinfo {author}
			{\bibfnamefont {C.}~\bibnamefont {Hern{\'{a}}ndez-Garc{\'{i}}a}}, \bibinfo
			{author} {\bibfnamefont {D.~D.}\ \bibnamefont {Hickstein}}, \bibinfo {author}
			{\bibfnamefont {D.}~\bibnamefont {Zusin}}, \bibinfo {author} {\bibfnamefont
				{C.}~\bibnamefont {Gentry}}, \bibinfo {author} {\bibfnamefont {F.~J.}\
				\bibnamefont {Dollar}}, \bibinfo {author} {\bibfnamefont {C.~A.}\
				\bibnamefont {Mancuso}}, \bibinfo {author} {\bibfnamefont {C.~W.}\
				\bibnamefont {Hogle}}, \bibinfo {author} {\bibfnamefont {O.}~\bibnamefont
				{Kfir}}, \bibinfo {author} {\bibfnamefont {D.}~\bibnamefont {Legut}},
			\bibinfo {author} {\bibfnamefont {K.}~\bibnamefont {Carva}}, \bibinfo
			{author} {\bibfnamefont {J.~L.}\ \bibnamefont {Ellis}}, \bibinfo {author}
			{\bibfnamefont {K.~M.}\ \bibnamefont {Dorney}}, \bibinfo {author}
			{\bibfnamefont {C.}~\bibnamefont {Chen}}, \bibinfo {author} {\bibfnamefont
				{O.~G.}\ \bibnamefont {Shpyrko}}, \bibinfo {author} {\bibfnamefont {E.~E.}\
				\bibnamefont {Fullerton}}, \bibinfo {author} {\bibfnamefont {O.}~\bibnamefont
				{Cohen}}, \bibinfo {author} {\bibfnamefont {P.~M.}\ \bibnamefont {Oppeneer}},
			\bibinfo {author} {\bibfnamefont {D.~B.}\ \bibnamefont
				{Milo{\v{s}}evi{\'{c}}}}, \bibinfo {author} {\bibfnamefont {A.}~\bibnamefont
				{Becker}}, \bibinfo {author} {\bibfnamefont {A.~A.}\ \bibnamefont
				{Jaro{\'{n}}-Becker}}, \bibinfo {author} {\bibfnamefont {T.}~\bibnamefont
				{Popmintchev}}, \bibinfo {author} {\bibfnamefont {M.~M.}\ \bibnamefont
				{Murnane}},\ and\ \bibinfo {author} {\bibfnamefont {H.~C.}\ \bibnamefont
				{Kapteyn}},\ }\bibfield  {title} {\bibinfo {title} {{Bright circularly
					polarized soft X-ray high harmonics for X-ray magnetic circular dichroism}},\
		}\href {https://doi.org/10.1073/pnas.1519666112} {\bibfield  {journal}
			{\bibinfo  {journal} {Proc. Natl. Acad. Sci.}\ }\textbf {\bibinfo {volume}
				{112}},\ \bibinfo {pages} {14206} (\bibinfo {year} {2015})}\BibitemShut
		{NoStop}%
		\bibitem [{\citenamefont {Vodungbo}\ \emph {et~al.}(2011)\citenamefont
			{Vodungbo}, \citenamefont {{Barszczak Sardinha}}, \citenamefont {Gautier},
			\citenamefont {Lambert}, \citenamefont {Valentin}, \citenamefont {Lozano},
			\citenamefont {Iaquaniello}, \citenamefont {Delmotte}, \citenamefont
			{Sebban}, \citenamefont {L{\"{u}}ning},\ and\ \citenamefont
			{Zeitoun}}]{Vodungbo2011}%
		\BibitemOpen
		\bibfield  {author} {\bibinfo {author} {\bibfnamefont {B.}~\bibnamefont
				{Vodungbo}}, \bibinfo {author} {\bibfnamefont {A.}~\bibnamefont {{Barszczak
						Sardinha}}}, \bibinfo {author} {\bibfnamefont {J.}~\bibnamefont {Gautier}},
			\bibinfo {author} {\bibfnamefont {G.}~\bibnamefont {Lambert}}, \bibinfo
			{author} {\bibfnamefont {C.}~\bibnamefont {Valentin}}, \bibinfo {author}
			{\bibfnamefont {M.}~\bibnamefont {Lozano}}, \bibinfo {author} {\bibfnamefont
				{G.}~\bibnamefont {Iaquaniello}}, \bibinfo {author} {\bibfnamefont
				{F.}~\bibnamefont {Delmotte}}, \bibinfo {author} {\bibfnamefont
				{S.}~\bibnamefont {Sebban}}, \bibinfo {author} {\bibfnamefont
				{J.}~\bibnamefont {L{\"{u}}ning}},\ and\ \bibinfo {author} {\bibfnamefont
				{P.}~\bibnamefont {Zeitoun}},\ }\bibfield  {title} {\bibinfo {title}
			{{Polarization control of high order harmonics in the EUV photon energy
					range}},\ }\href {https://doi.org/10.1364/OE.19.004346} {\bibfield  {journal}
			{\bibinfo  {journal} {Opt. Express}\ }\textbf {\bibinfo {volume} {19}},\
			\bibinfo {pages} {4346} (\bibinfo {year} {2011})}\BibitemShut {NoStop}%
		\bibitem [{\citenamefont {Yao}\ \emph {et~al.}(2020)\citenamefont {Yao},
			\citenamefont {Willems}, \citenamefont {{von Korff Schmising}}, \citenamefont
			{Str{\"u}ber}, \citenamefont {Hessing}, \citenamefont {Pfau}, \citenamefont
			{Schick}, \citenamefont {Engel}, \citenamefont {Gerlinger}, \citenamefont
			{Schneider},\ and\ \citenamefont {Eisebitt}}]{Yao2020}%
		\BibitemOpen
		\bibfield  {author} {\bibinfo {author} {\bibfnamefont {K.}~\bibnamefont
				{Yao}}, \bibinfo {author} {\bibfnamefont {F.}~\bibnamefont {Willems}},
			\bibinfo {author} {\bibfnamefont {C.}~\bibnamefont {{von Korff Schmising}}},
			\bibinfo {author} {\bibfnamefont {C.}~\bibnamefont {Str{\"u}ber}}, \bibinfo
			{author} {\bibfnamefont {P.}~\bibnamefont {Hessing}}, \bibinfo {author}
			{\bibfnamefont {B.}~\bibnamefont {Pfau}}, \bibinfo {author} {\bibfnamefont
				{D.}~\bibnamefont {Schick}}, \bibinfo {author} {\bibfnamefont
				{D.}~\bibnamefont {Engel}}, \bibinfo {author} {\bibfnamefont
				{K.}~\bibnamefont {Gerlinger}}, \bibinfo {author} {\bibfnamefont
				{M.}~\bibnamefont {Schneider}},\ and\ \bibinfo {author} {\bibfnamefont
				{S.}~\bibnamefont {Eisebitt}},\ }\bibfield  {title} {\bibinfo {title} {A
				tabletop setup for ultrafast helicity-dependent and element-specific
				absorption spectroscopy and scattering in the extreme ultraviolet spectral
				range},\ }\href {https://doi.org/10.1063/5.0013928} {\bibfield  {journal}
			{\bibinfo  {journal} {Review of Scientific Instruments}\ }\textbf {\bibinfo
				{volume} {91}},\ \bibinfo {pages} {093001} (\bibinfo {year}
			{2020})}\BibitemShut {NoStop}%
		\bibitem [{\citenamefont {Feng}\ \emph {et~al.}(2020)\citenamefont {Feng},
			\citenamefont {Feng}, \citenamefont {Feng}, \citenamefont {Heilmann},
			\citenamefont {Heilmann}, \citenamefont {Bock}, \citenamefont {Ehrentraut},
			\citenamefont {Witting}, \citenamefont {Yu}, \citenamefont {Stiel},
			\citenamefont {Eisebitt},\ and\ \citenamefont {Schn{\"u}rer}}]{Feng2020a}%
		\BibitemOpen
		\bibfield  {author} {\bibinfo {author} {\bibfnamefont {T.}~\bibnamefont
				{Feng}}, \bibinfo {author} {\bibfnamefont {T.}~\bibnamefont {Feng}}, \bibinfo
			{author} {\bibfnamefont {T.}~\bibnamefont {Feng}}, \bibinfo {author}
			{\bibfnamefont {A.}~\bibnamefont {Heilmann}}, \bibinfo {author}
			{\bibfnamefont {A.}~\bibnamefont {Heilmann}}, \bibinfo {author}
			{\bibfnamefont {M.}~\bibnamefont {Bock}}, \bibinfo {author} {\bibfnamefont
				{L.}~\bibnamefont {Ehrentraut}}, \bibinfo {author} {\bibfnamefont
				{T.}~\bibnamefont {Witting}}, \bibinfo {author} {\bibfnamefont
				{H.}~\bibnamefont {Yu}}, \bibinfo {author} {\bibfnamefont {H.}~\bibnamefont
				{Stiel}}, \bibinfo {author} {\bibfnamefont {S.}~\bibnamefont {Eisebitt}},\
			and\ \bibinfo {author} {\bibfnamefont {M.}~\bibnamefont {Schn{\"u}rer}},\
		}\bibfield  {title} {\bibinfo {title} {27 w 2.1 \textmu m opcpa system for
				coherent soft x-ray generation operating at 10 khz},\ }\href
		{https://doi.org/10.1364/OE.386588} {\bibfield  {journal} {\bibinfo
				{journal} {Opt. Express, OE}\ }\textbf {\bibinfo {volume} {28}},\ \bibinfo
			{pages} {8724} (\bibinfo {year} {2020})}\BibitemShut {NoStop}%
		\bibitem [{\citenamefont {Schick}\ \emph {et~al.}(2021)\citenamefont {Schick},
			\citenamefont {Borchert}, \citenamefont {Braenzel}, \citenamefont {Stiel},
			\citenamefont {Tümmler}, \citenamefont {Bürgler}, \citenamefont {Firsov},
			\citenamefont {Schmising}, \citenamefont {Pfau}, \citenamefont {Eisebitt},\
			and\ \citenamefont {Eisebitt}}]{schick_laser-driven_2021}%
		\BibitemOpen
		\bibfield  {author} {\bibinfo {author} {\bibfnamefont {D.}~\bibnamefont
				{Schick}}, \bibinfo {author} {\bibfnamefont {M.}~\bibnamefont {Borchert}},
			\bibinfo {author} {\bibfnamefont {J.}~\bibnamefont {Braenzel}}, \bibinfo
			{author} {\bibfnamefont {H.}~\bibnamefont {Stiel}}, \bibinfo {author}
			{\bibfnamefont {J.}~\bibnamefont {Tümmler}}, \bibinfo {author}
			{\bibfnamefont {D.~E.}\ \bibnamefont {Bürgler}}, \bibinfo {author}
			{\bibfnamefont {A.}~\bibnamefont {Firsov}}, \bibinfo {author} {\bibfnamefont
				{C.~v.~K.}\ \bibnamefont {Schmising}}, \bibinfo {author} {\bibfnamefont
				{B.}~\bibnamefont {Pfau}}, \bibinfo {author} {\bibfnamefont {S.}~\bibnamefont
				{Eisebitt}},\ and\ \bibinfo {author} {\bibfnamefont {S.}~\bibnamefont
				{Eisebitt}},\ }\bibfield  {title} {\bibinfo {title} {Laser-driven resonant
				magnetic soft-x-ray scattering for probing ultrafast antiferromagnetic and
				structural dynamics},\ }\href {https://doi.org/10.1364/OPTICA.435522}
		{\bibfield  {journal} {\bibinfo  {journal} {Optica, OPTICA}\ }\textbf
			{\bibinfo {volume} {8}},\ \bibinfo {pages} {1237} (\bibinfo {year} {2021})},\
		\bibinfo {note} {publisher: Optica Publishing Group}\BibitemShut {NoStop}%
		\bibitem [{\citenamefont {Kortright}\ \emph {et~al.}(1997)\citenamefont
			{Kortright}, \citenamefont {Kim}, \citenamefont {Warwick},\ and\
			\citenamefont {Smith}}]{Kortright1997}%
		\BibitemOpen
		\bibfield  {author} {\bibinfo {author} {\bibfnamefont {J.~B.}\ \bibnamefont
				{Kortright}}, \bibinfo {author} {\bibfnamefont {S.-K.}\ \bibnamefont {Kim}},
			\bibinfo {author} {\bibfnamefont {T.}~\bibnamefont {Warwick}},\ and\ \bibinfo
			{author} {\bibfnamefont {N.~V.}\ \bibnamefont {Smith}},\ }\bibfield  {title}
		{\bibinfo {title} {{Soft x-ray circular polarizer using magnetic circular
					dichroism at the Fe L3 line}},\ }\href {https://doi.org/10.1063/1.119932}
		{\bibfield  {journal} {\bibinfo  {journal} {Appl. Phys. Lett.}\ }\textbf
			{\bibinfo {volume} {71}},\ \bibinfo {pages} {1446} (\bibinfo {year}
			{1997})}\BibitemShut {NoStop}%
		\bibitem [{\citenamefont {Pfau}\ \emph {et~al.}(2010)\citenamefont {Pfau},
			\citenamefont {G{\"{u}}nther}, \citenamefont {K{\"{o}}nnecke}, \citenamefont
			{Guehrs}, \citenamefont {Hellwig}, \citenamefont {Schlotter},\ and\
			\citenamefont {Eisebitt}}]{Pfau2010}%
		\BibitemOpen
		\bibfield  {author} {\bibinfo {author} {\bibfnamefont {B.}~\bibnamefont
				{Pfau}}, \bibinfo {author} {\bibfnamefont {C.~M.}\ \bibnamefont
				{G{\"{u}}nther}}, \bibinfo {author} {\bibfnamefont {R.}~\bibnamefont
				{K{\"{o}}nnecke}}, \bibinfo {author} {\bibfnamefont {E.}~\bibnamefont
				{Guehrs}}, \bibinfo {author} {\bibfnamefont {O.}~\bibnamefont {Hellwig}},
			\bibinfo {author} {\bibfnamefont {W.~F.}\ \bibnamefont {Schlotter}},\ and\
			\bibinfo {author} {\bibfnamefont {S.}~\bibnamefont {Eisebitt}},\ }\bibfield
		{title} {\bibinfo {title} {{Magnetic imaging at linearly polarized x-ray
					sources}},\ }\href {https://doi.org/10.1364/OE.18.013608} {\bibfield
			{journal} {\bibinfo  {journal} {Opt. Express}\ }\textbf {\bibinfo {volume}
				{18}},\ \bibinfo {pages} {13608} (\bibinfo {year} {2010})}\BibitemShut
		{NoStop}%
		\bibitem [{\citenamefont {Stanciu}\ \emph {et~al.}(2007)\citenamefont
			{Stanciu}, \citenamefont {Hansteen}, \citenamefont {Kimel}, \citenamefont
			{Kirilyuk}, \citenamefont {Tsukamoto}, \citenamefont {Itoh},\ and\
			\citenamefont {Rasing}}]{Stanciu2007}%
		\BibitemOpen
		\bibfield  {author} {\bibinfo {author} {\bibfnamefont {C.~D.}\ \bibnamefont
				{Stanciu}}, \bibinfo {author} {\bibfnamefont {F.}~\bibnamefont {Hansteen}},
			\bibinfo {author} {\bibfnamefont {A.~V.}\ \bibnamefont {Kimel}}, \bibinfo
			{author} {\bibfnamefont {A.}~\bibnamefont {Kirilyuk}}, \bibinfo {author}
			{\bibfnamefont {A.}~\bibnamefont {Tsukamoto}}, \bibinfo {author}
			{\bibfnamefont {A.}~\bibnamefont {Itoh}},\ and\ \bibinfo {author}
			{\bibfnamefont {T.}~\bibnamefont {Rasing}},\ }\bibfield  {title} {\bibinfo
			{title} {All-optical magnetic recording with circularly polarized light},\
		}\href {https://doi.org/10.1103/PhysRevLett.99.047601} {\bibfield  {journal}
			{\bibinfo  {journal} {Phys. Rev. Lett.}\ }\textbf {\bibinfo {volume} {99}},\
			\bibinfo {pages} {047601} (\bibinfo {year} {2007})}\BibitemShut {NoStop}%
		\bibitem [{\citenamefont {Li}\ \emph {et~al.}(2022)\citenamefont {Li},
			\citenamefont {Medapalli}, \citenamefont {Mentink}, \citenamefont
			{Mikhaylovskiy}, \citenamefont {Blank}, \citenamefont {Patel}, \citenamefont
			{Zvezdin}, \citenamefont {Rasing}, \citenamefont {Fullerton},\ and\
			\citenamefont {Kimel}}]{Li2022}%
		\BibitemOpen
		\bibfield  {author} {\bibinfo {author} {\bibfnamefont {G.}~\bibnamefont
				{Li}}, \bibinfo {author} {\bibfnamefont {R.}~\bibnamefont {Medapalli}},
			\bibinfo {author} {\bibfnamefont {J.~H.}\ \bibnamefont {Mentink}}, \bibinfo
			{author} {\bibfnamefont {R.~V.}\ \bibnamefont {Mikhaylovskiy}}, \bibinfo
			{author} {\bibfnamefont {T.~G.~H.}\ \bibnamefont {Blank}}, \bibinfo {author}
			{\bibfnamefont {S.~K.~K.}\ \bibnamefont {Patel}}, \bibinfo {author}
			{\bibfnamefont {A.~K.}\ \bibnamefont {Zvezdin}}, \bibinfo {author}
			{\bibfnamefont {T.}~\bibnamefont {Rasing}}, \bibinfo {author} {\bibfnamefont
				{E.~E.}\ \bibnamefont {Fullerton}},\ and\ \bibinfo {author} {\bibfnamefont
				{A.~V.}\ \bibnamefont {Kimel}},\ }\bibfield  {title} {\bibinfo {title}
			{Ultrafast kinetics of the antiferromagnetic-ferromagnetic phase transition
				in ferh},\ }\href {https://doi.org/10.1038/s41467-022-30591-2} {\bibfield
			{journal} {\bibinfo  {journal} {Nat Commun}\ }\textbf {\bibinfo {volume}
				{13}},\ \bibinfo {pages} {2998} (\bibinfo {year} {2022})}\BibitemShut
		{NoStop}%
		\bibitem [{\citenamefont {Jung}\ \emph {et~al.}(2016)\citenamefont {Jung},
			\citenamefont {T{\"{u}}mmler},\ and\ \citenamefont {Will}}]{Jung2016}%
		\BibitemOpen
		\bibfield  {author} {\bibinfo {author} {\bibfnamefont {R.}~\bibnamefont
				{Jung}}, \bibinfo {author} {\bibfnamefont {J.}~\bibnamefont
				{T{\"{u}}mmler}},\ and\ \bibinfo {author} {\bibfnamefont {I.}~\bibnamefont
				{Will}},\ }\bibfield  {title} {\bibinfo {title} {{Regenerative thin-disk
					amplifier for 300 mJ pulse energy}},\ }\href
		{https://doi.org/10.1364/OE.24.000883} {\bibfield  {journal} {\bibinfo
				{journal} {Opt. Express}\ }\textbf {\bibinfo {volume} {24}},\ \bibinfo
			{pages} {883} (\bibinfo {year} {2016})}\BibitemShut {NoStop}%
		\bibitem [{\citenamefont {Mantouvalou}\ \emph {et~al.}(2015)\citenamefont
			{Mantouvalou}, \citenamefont {Witte}, \citenamefont {Grötzsch},
			\citenamefont {Neitzel}, \citenamefont {Günther}, \citenamefont {Baumann},
			\citenamefont {Jung}, \citenamefont {Stiel}, \citenamefont {Kanngießer},\
			and\ \citenamefont {Sandner}}]{mantouvalou_high_2015}%
		\BibitemOpen
		\bibfield  {author} {\bibinfo {author} {\bibfnamefont {I.}~\bibnamefont
				{Mantouvalou}}, \bibinfo {author} {\bibfnamefont {K.}~\bibnamefont {Witte}},
			\bibinfo {author} {\bibfnamefont {D.}~\bibnamefont {Grötzsch}}, \bibinfo
			{author} {\bibfnamefont {M.}~\bibnamefont {Neitzel}}, \bibinfo {author}
			{\bibfnamefont {S.}~\bibnamefont {Günther}}, \bibinfo {author}
			{\bibfnamefont {J.}~\bibnamefont {Baumann}}, \bibinfo {author} {\bibfnamefont
				{R.}~\bibnamefont {Jung}}, \bibinfo {author} {\bibfnamefont {H.}~\bibnamefont
				{Stiel}}, \bibinfo {author} {\bibfnamefont {B.}~\bibnamefont {Kanngießer}},\
			and\ \bibinfo {author} {\bibfnamefont {W.}~\bibnamefont {Sandner}},\
		}\bibfield  {title} {\bibinfo {title} {High average power, highly brilliant
				laser-produced plasma source for soft {X}-ray spectroscopy},\ }\href
		{https://doi.org/10.1063/1.4916193} {\bibfield  {journal} {\bibinfo
				{journal} {RSI}\ }\textbf {\bibinfo {volume} {86}},\ \bibinfo {pages}
			{035116} (\bibinfo {year} {2015})},\ \bibinfo {note} {publisher: American
			Institute of Physics}\BibitemShut {NoStop}%
		\bibitem [{\citenamefont {Brzhezinskaya}\ \emph {et~al.}(2013)\citenamefont
			{Brzhezinskaya}, \citenamefont {Firsov}, \citenamefont {Holldack},
			\citenamefont {Kachel}, \citenamefont {Mitzner}, \citenamefont {Pontius},
			\citenamefont {Schmidt}, \citenamefont {Sperling}, \citenamefont {Stamm},
			\citenamefont {F{\"{o}}hlisch},\ and\ \citenamefont
			{Erko}}]{Brzhezinskaya2013}%
		\BibitemOpen
		\bibfield  {author} {\bibinfo {author} {\bibfnamefont {M.}~\bibnamefont
				{Brzhezinskaya}}, \bibinfo {author} {\bibfnamefont {A.}~\bibnamefont
				{Firsov}}, \bibinfo {author} {\bibfnamefont {K.}~\bibnamefont {Holldack}},
			\bibinfo {author} {\bibfnamefont {T.}~\bibnamefont {Kachel}}, \bibinfo
			{author} {\bibfnamefont {R.}~\bibnamefont {Mitzner}}, \bibinfo {author}
			{\bibfnamefont {N.}~\bibnamefont {Pontius}}, \bibinfo {author} {\bibfnamefont
				{J.-S.}\ \bibnamefont {Schmidt}}, \bibinfo {author} {\bibfnamefont
				{M.}~\bibnamefont {Sperling}}, \bibinfo {author} {\bibfnamefont
				{C.}~\bibnamefont {Stamm}}, \bibinfo {author} {\bibfnamefont
				{A.}~\bibnamefont {F{\"{o}}hlisch}},\ and\ \bibinfo {author} {\bibfnamefont
				{A.}~\bibnamefont {Erko}},\ }\bibfield  {title} {\bibinfo {title} {{A novel
					monochromator for experiments with ultrashort X-ray pulses}},\ }\href
		{https://doi.org/10.1107/S0909049513008613} {\bibfield  {journal} {\bibinfo
				{journal} {J. Synchrotron Radiat.}\ }\textbf {\bibinfo {volume} {20}},\
			\bibinfo {pages} {522} (\bibinfo {year} {2013})}\BibitemShut {NoStop}%
		\bibitem [{\citenamefont {Dirac}(1930)}]{Dirac1930}%
		\BibitemOpen
		\bibfield  {author} {\bibinfo {author} {\bibfnamefont {P.~A.~M.}\
				\bibnamefont {Dirac}},\ }\href@noop {} {\emph {\bibinfo {title} {{The
						principles of quantum mechanics}}}},\ \bibinfo {edition} {4th}\ ed.\
		(\bibinfo  {publisher} {Oxford University Press},\ \bibinfo {year}
		{1930})\BibitemShut {NoStop}%
		\bibitem [{\citenamefont {Legall}\ \emph {et~al.}(2004)\citenamefont {Legall},
			\citenamefont {Stiel}, \citenamefont {Vogt}, \citenamefont {Schoennagel},
			\citenamefont {Nickles}, \citenamefont {Tuemmler}, \citenamefont {Scholz},\
			and\ \citenamefont {Scholze}}]{legall_spatial_2004}%
		\BibitemOpen
		\bibfield  {author} {\bibinfo {author} {\bibfnamefont {H.}~\bibnamefont
				{Legall}}, \bibinfo {author} {\bibfnamefont {H.}~\bibnamefont {Stiel}},
			\bibinfo {author} {\bibfnamefont {U.}~\bibnamefont {Vogt}}, \bibinfo {author}
			{\bibfnamefont {H.}~\bibnamefont {Schoennagel}}, \bibinfo {author}
			{\bibfnamefont {P.-V.}\ \bibnamefont {Nickles}}, \bibinfo {author}
			{\bibfnamefont {J.}~\bibnamefont {Tuemmler}}, \bibinfo {author}
			{\bibfnamefont {F.}~\bibnamefont {Scholz}},\ and\ \bibinfo {author}
			{\bibfnamefont {F.}~\bibnamefont {Scholze}},\ }\bibfield  {title}
		{{\selectlanguage {English}\bibinfo {title} {Spatial and spectral
					characterization of a laser produced plasma source for extreme ultraviolet
					metrology}},\ }\bibfield  {journal} {\bibinfo  {journal} {Review of
				Scientific Instruments}\ }\textbf {\bibinfo {volume} {75}},\ \href
		{https://doi.org/10.1063/1.1807567} {10.1063/1.1807567} (\bibinfo {year}
		{2004})\BibitemShut {NoStop}%
		\bibitem [{\citenamefont {Schick}(2021)}]{Schick2021b}%
		\BibitemOpen
		\bibfield  {author} {\bibinfo {author} {\bibfnamefont {D.}~\bibnamefont
				{Schick}},\ }\bibfield  {title} {\bibinfo {title} {{udkm1Dsim – a Python
					toolbox for simulating 1D ultrafast dynamics in condensed matter}},\ }\href
		{https://doi.org/10.1016/j.cpc.2021.108031} {\bibfield  {journal} {\bibinfo
				{journal} {Comput. Phys. Commun.}\ }\textbf {\bibinfo {volume} {266}},\
			\bibinfo {pages} {108031} (\bibinfo {year} {2021})},\ \Eprint
		{https://arxiv.org/abs/2102.12144} {2102.12144} \BibitemShut {NoStop}%
		\bibitem [{\citenamefont {Elzo}\ \emph {et~al.}(2012)\citenamefont {Elzo},
			\citenamefont {Jal}, \citenamefont {Bunau}, \citenamefont {Grenier},
			\citenamefont {Joly}, \citenamefont {Ramos}, \citenamefont {Tolentino},
			\citenamefont {Tonnerre},\ and\ \citenamefont {Jaouen}}]{Elzo2012}%
		\BibitemOpen
		\bibfield  {author} {\bibinfo {author} {\bibfnamefont {M.}~\bibnamefont
				{Elzo}}, \bibinfo {author} {\bibfnamefont {E.}~\bibnamefont {Jal}}, \bibinfo
			{author} {\bibfnamefont {O.}~\bibnamefont {Bunau}}, \bibinfo {author}
			{\bibfnamefont {S.}~\bibnamefont {Grenier}}, \bibinfo {author} {\bibfnamefont
				{Y.}~\bibnamefont {Joly}}, \bibinfo {author} {\bibfnamefont {A.}~\bibnamefont
				{Ramos}}, \bibinfo {author} {\bibfnamefont {H.}~\bibnamefont {Tolentino}},
			\bibinfo {author} {\bibfnamefont {J.}~\bibnamefont {Tonnerre}},\ and\
			\bibinfo {author} {\bibfnamefont {N.}~\bibnamefont {Jaouen}},\ }\bibfield
		{title} {\bibinfo {title} {{X-ray resonant magnetic reflectivity of
					stratified magnetic structures: Eigenwave formalism and application to a
					W/Fe/W trilayer}},\ }\href {https://doi.org/10.1016/j.jmmm.2011.07.019}
		{\bibfield  {journal} {\bibinfo  {journal} {J. Magn. Magn. Mater.}\ }\textbf
			{\bibinfo {volume} {324}},\ \bibinfo {pages} {105} (\bibinfo {year}
			{2012})}\BibitemShut {NoStop}%
		\bibitem [{\citenamefont {Chen}\ \emph {et~al.}(1995)\citenamefont {Chen},
			\citenamefont {Idzerda}, \citenamefont {Lin}, \citenamefont {Smith},
			\citenamefont {Meigs}, \citenamefont {Chaban}, \citenamefont {Ho},
			\citenamefont {Pellegrin},\ and\ \citenamefont {Sette}}]{Chen1995}%
		\BibitemOpen
		\bibfield  {author} {\bibinfo {author} {\bibfnamefont {C.~T.}\ \bibnamefont
				{Chen}}, \bibinfo {author} {\bibfnamefont {Y.~U.}\ \bibnamefont {Idzerda}},
			\bibinfo {author} {\bibfnamefont {H.-J.}\ \bibnamefont {Lin}}, \bibinfo
			{author} {\bibfnamefont {N.~V.}\ \bibnamefont {Smith}}, \bibinfo {author}
			{\bibfnamefont {G.}~\bibnamefont {Meigs}}, \bibinfo {author} {\bibfnamefont
				{E.}~\bibnamefont {Chaban}}, \bibinfo {author} {\bibfnamefont {G.~H.}\
				\bibnamefont {Ho}}, \bibinfo {author} {\bibfnamefont {E.}~\bibnamefont
				{Pellegrin}},\ and\ \bibinfo {author} {\bibfnamefont {F.}~\bibnamefont
				{Sette}},\ }\bibfield  {title} {\bibinfo {title} {{Experimental Confirmation
					of the X-Ray Magnetic Circular Dichroism Sum Rules for Iron and Cobalt}},\
		}\href {https://doi.org/10.1103/PhysRevLett.75.152} {\bibfield  {journal}
			{\bibinfo  {journal} {Phys. Rev. Lett.}\ }\textbf {\bibinfo {volume} {75}},\
			\bibinfo {pages} {152} (\bibinfo {year} {1995})}\BibitemShut {NoStop}%
		\bibitem [{\citenamefont {Peters}\ \emph {et~al.}(2004)\citenamefont {Peters},
			\citenamefont {Miguel}, \citenamefont {{de Vries}}, \citenamefont
			{Toulemonde}, \citenamefont {Goedkoop}, \citenamefont {Dhesi},\ and\
			\citenamefont {Brookes}}]{Peters2004}%
		\BibitemOpen
		\bibfield  {author} {\bibinfo {author} {\bibfnamefont {J.~F.}\ \bibnamefont
				{Peters}}, \bibinfo {author} {\bibfnamefont {J.}~\bibnamefont {Miguel}},
			\bibinfo {author} {\bibfnamefont {M.~A.}\ \bibnamefont {{de Vries}}},
			\bibinfo {author} {\bibfnamefont {O.~M.}\ \bibnamefont {Toulemonde}},
			\bibinfo {author} {\bibfnamefont {J.~B.}\ \bibnamefont {Goedkoop}}, \bibinfo
			{author} {\bibfnamefont {S.~S.}\ \bibnamefont {Dhesi}},\ and\ \bibinfo
			{author} {\bibfnamefont {N.~B.}\ \bibnamefont {Brookes}},\ }\bibfield
		{title} {\bibinfo {title} {Soft x-ray resonant magneto-optical constants at
				the gd ${M}_{4,5}$ and fe ${L}_{2,3}$ edges},\ }\href
		{https://doi.org/10.1103/PhysRevB.70.224417} {\bibfield  {journal} {\bibinfo
				{journal} {Phys. Rev. B}\ }\textbf {\bibinfo {volume} {70}},\ \bibinfo
			{pages} {224417} (\bibinfo {year} {2004})}\BibitemShut {NoStop}%
		\bibitem [{\citenamefont {Bergamaschi}\ \emph {et~al.}(2018)\citenamefont
			{Bergamaschi}, \citenamefont {Andr{\"a}}, \citenamefont {Barten},
			\citenamefont {Borca}, \citenamefont {Br{\"u}ckner}, \citenamefont
			{Chiriotti}, \citenamefont {Dinapoli}, \citenamefont {Fr{\"o}jdh},
			\citenamefont {Greiffenberg}, \citenamefont {Huthwelker}, \citenamefont
			{Kleibert}, \citenamefont {Langer}, \citenamefont {Lebugle}, \citenamefont
			{{Lopez-Cuenca}}, \citenamefont {Mezza}, \citenamefont {Mozzanica},
			\citenamefont {Raabe}, \citenamefont {Redford}, \citenamefont {Ruder},
			\citenamefont {Scagnoli}, \citenamefont {Schmitt}, \citenamefont {Shi},
			\citenamefont {Staub}, \citenamefont {Thattil}, \citenamefont {Tinti},
			\citenamefont {Vaz}, \citenamefont {Vetter}, \citenamefont
			{{Vila-Comamala}},\ and\ \citenamefont {Zhang}}]{Bergamaschi2018a}%
		\BibitemOpen
		\bibfield  {author} {\bibinfo {author} {\bibfnamefont {A.}~\bibnamefont
				{Bergamaschi}}, \bibinfo {author} {\bibfnamefont {M.}~\bibnamefont
				{Andr{\"a}}}, \bibinfo {author} {\bibfnamefont {R.}~\bibnamefont {Barten}},
			\bibinfo {author} {\bibfnamefont {C.}~\bibnamefont {Borca}}, \bibinfo
			{author} {\bibfnamefont {M.}~\bibnamefont {Br{\"u}ckner}}, \bibinfo {author}
			{\bibfnamefont {S.}~\bibnamefont {Chiriotti}}, \bibinfo {author}
			{\bibfnamefont {R.}~\bibnamefont {Dinapoli}}, \bibinfo {author}
			{\bibfnamefont {E.}~\bibnamefont {Fr{\"o}jdh}}, \bibinfo {author}
			{\bibfnamefont {D.}~\bibnamefont {Greiffenberg}}, \bibinfo {author}
			{\bibfnamefont {T.}~\bibnamefont {Huthwelker}}, \bibinfo {author}
			{\bibfnamefont {A.}~\bibnamefont {Kleibert}}, \bibinfo {author}
			{\bibfnamefont {M.}~\bibnamefont {Langer}}, \bibinfo {author} {\bibfnamefont
				{M.}~\bibnamefont {Lebugle}}, \bibinfo {author} {\bibfnamefont
				{C.}~\bibnamefont {{Lopez-Cuenca}}}, \bibinfo {author} {\bibfnamefont
				{D.}~\bibnamefont {Mezza}}, \bibinfo {author} {\bibfnamefont
				{A.}~\bibnamefont {Mozzanica}}, \bibinfo {author} {\bibfnamefont
				{J.}~\bibnamefont {Raabe}}, \bibinfo {author} {\bibfnamefont
				{S.}~\bibnamefont {Redford}}, \bibinfo {author} {\bibfnamefont
				{C.}~\bibnamefont {Ruder}}, \bibinfo {author} {\bibfnamefont
				{V.}~\bibnamefont {Scagnoli}}, \bibinfo {author} {\bibfnamefont
				{B.}~\bibnamefont {Schmitt}}, \bibinfo {author} {\bibfnamefont
				{X.}~\bibnamefont {Shi}}, \bibinfo {author} {\bibfnamefont {U.}~\bibnamefont
				{Staub}}, \bibinfo {author} {\bibfnamefont {D.}~\bibnamefont {Thattil}},
			\bibinfo {author} {\bibfnamefont {G.}~\bibnamefont {Tinti}}, \bibinfo
			{author} {\bibfnamefont {C.~F.}\ \bibnamefont {Vaz}}, \bibinfo {author}
			{\bibfnamefont {S.}~\bibnamefont {Vetter}}, \bibinfo {author} {\bibfnamefont
				{J.}~\bibnamefont {{Vila-Comamala}}},\ and\ \bibinfo {author} {\bibfnamefont
				{J.}~\bibnamefont {Zhang}},\ }\bibfield  {title} {\bibinfo {title} {The
				m\"onch detector for soft x-ray, high-resolution, and energy resolved
				applications},\ }\href {https://doi.org/10.1080/08940886.2018.1528428}
		{\bibfield  {journal} {\bibinfo  {journal} {Synchrotron Radiation News}\
			}\textbf {\bibinfo {volume} {31}},\ \bibinfo {pages} {11} (\bibinfo {year}
			{2018})}\BibitemShut {NoStop}%
		\bibitem [{\citenamefont {Dullin}\ \emph {et~al.}(2018)\citenamefont {Dullin},
			\citenamefont {Albers}, \citenamefont {Tromba}, \citenamefont {Andr{\"a}},
			\citenamefont {Ramilli},\ and\ \citenamefont {Bergamaschi}}]{Dullin2018}%
		\BibitemOpen
		\bibfield  {author} {\bibinfo {author} {\bibfnamefont {C.}~\bibnamefont
				{Dullin}}, \bibinfo {author} {\bibfnamefont {J.}~\bibnamefont {Albers}},
			\bibinfo {author} {\bibfnamefont {G.}~\bibnamefont {Tromba}}, \bibinfo
			{author} {\bibfnamefont {M.}~\bibnamefont {Andr{\"a}}}, \bibinfo {author}
			{\bibfnamefont {M.}~\bibnamefont {Ramilli}},\ and\ \bibinfo {author}
			{\bibfnamefont {A.}~\bibnamefont {Bergamaschi}},\ }\bibfield  {title}
		{\bibinfo {title} {M\"onch detector enables fast and low-dose
				free-propagation phase-contrast computed tomography of in situ mouse lungs},\
		}\href {https://doi.org/10.1107/S160057751701668X} {\bibfield  {journal}
			{\bibinfo  {journal} {J Synchrotron Rad}\ }\textbf {\bibinfo {volume} {25}},\
			\bibinfo {pages} {565} (\bibinfo {year} {2018})}\BibitemShut {NoStop}%
		\bibitem [{\citenamefont {Pfaff}\ \emph {et~al.}(2021)\citenamefont {Pfaff},
			\citenamefont {Rampp}, \citenamefont {Herkommer}, \citenamefont {Jung},
			\citenamefont {Teisset}, \citenamefont {Klingebiel},\ and\ \citenamefont
			{Metzger}}]{Pfaff2021}%
		\BibitemOpen
		\bibfield  {author} {\bibinfo {author} {\bibfnamefont {Y.}~\bibnamefont
				{Pfaff}}, \bibinfo {author} {\bibfnamefont {M.}~\bibnamefont {Rampp}},
			\bibinfo {author} {\bibfnamefont {C.}~\bibnamefont {Herkommer}}, \bibinfo
			{author} {\bibfnamefont {R.}~\bibnamefont {Jung}}, \bibinfo {author}
			{\bibfnamefont {C.~Y.}\ \bibnamefont {Teisset}}, \bibinfo {author}
			{\bibfnamefont {S.}~\bibnamefont {Klingebiel}},\ and\ \bibinfo {author}
			{\bibfnamefont {T.}~\bibnamefont {Metzger}},\ }\bibfield  {title} {\bibinfo
			{title} {Thin-disk based regenerative chirped pulse amplifier with 550 mj
				pulse energy at 1 khz repetition rate},\ }in\ \href
		{https://doi.org/10.1364/ASSL.2021.AM2A.5} {\emph {\bibinfo {booktitle}
				{Laser Congress 2021 (ASSL,LAC) (2021), paper AM2A.5}}}\ (\bibinfo
		{publisher} {Optica Publishing Group},\ \bibinfo {year} {2021})\ p.\ \bibinfo
		{pages} {AM2A.5}\BibitemShut {NoStop}%
		\bibitem [{\citenamefont {Battiato}\ \emph {et~al.}(2010)\citenamefont
			{Battiato}, \citenamefont {Carva},\ and\ \citenamefont
			{Oppeneer}}]{Battiato2010}%
		\BibitemOpen
		\bibfield  {author} {\bibinfo {author} {\bibfnamefont {M.}~\bibnamefont
				{Battiato}}, \bibinfo {author} {\bibfnamefont {K.}~\bibnamefont {Carva}},\
			and\ \bibinfo {author} {\bibfnamefont {P.~M.}\ \bibnamefont {Oppeneer}},\
		}\bibfield  {title} {\bibinfo {title} {Superdiffusive {Spin} {Transport} as a
				{Mechanism} of {Ultrafast} {Demagnetization}},\ }\href
		{https://doi.org/10.1103/PhysRevLett.105.027203} {\bibfield  {journal}
			{\bibinfo  {journal} {Phys. Rev. Lett.}\ }\textbf {\bibinfo {volume} {105}},\
			\bibinfo {pages} {027203} (\bibinfo {year} {2010})}\BibitemShut {NoStop}%
		\bibitem [{\citenamefont {Dewhurst}\ \emph {et~al.}(2018)\citenamefont
			{Dewhurst}, \citenamefont {Elliott}, \citenamefont {Shallcross},
			\citenamefont {Gross},\ and\ \citenamefont {Sharma}}]{Dewhurst2018}%
		\BibitemOpen
		\bibfield  {author} {\bibinfo {author} {\bibfnamefont {J.~K.}\ \bibnamefont
				{Dewhurst}}, \bibinfo {author} {\bibfnamefont {P.}~\bibnamefont {Elliott}},
			\bibinfo {author} {\bibfnamefont {S.}~\bibnamefont {Shallcross}}, \bibinfo
			{author} {\bibfnamefont {E.~K.~U.}\ \bibnamefont {Gross}},\ and\ \bibinfo
			{author} {\bibfnamefont {S.}~\bibnamefont {Sharma}},\ }\bibfield  {title}
		{\bibinfo {title} {Laser-{Induced} {Intersite} {Spin} {Transfer}},\ }\href
		{https://doi.org/10.1021/acs.nanolett.7b05118} {\bibfield  {journal}
			{\bibinfo  {journal} {Nano Lett.}\ }\textbf {\bibinfo {volume} {18}},\
			\bibinfo {pages} {1842} (\bibinfo {year} {2018})}\BibitemShut {NoStop}%
		\bibitem [{\citenamefont {Beens}\ \emph {et~al.}(2020)\citenamefont {Beens},
			\citenamefont {Duine},\ and\ \citenamefont {Koopmans}}]{Beens2020}%
		\BibitemOpen
		\bibfield  {author} {\bibinfo {author} {\bibfnamefont {M.}~\bibnamefont
				{Beens}}, \bibinfo {author} {\bibfnamefont {R.~A.}\ \bibnamefont {Duine}},\
			and\ \bibinfo {author} {\bibfnamefont {B.}~\bibnamefont {Koopmans}},\
		}\bibfield  {title} {\bibinfo {title} {$s-d$ model for local and nonlocal
				spin dynamics in laser-excited magnetic heterostructures},\ }\href
		{https://doi.org/10.1103/PhysRevB.102.054442} {\bibfield  {journal} {\bibinfo
				{journal} {Phys. Rev. B}\ }\textbf {\bibinfo {volume} {102}},\ \bibinfo
			{pages} {054442} (\bibinfo {year} {2020})}\BibitemShut {NoStop}%
		\bibitem [{\citenamefont {Burn}\ \emph {et~al.}(2020)\citenamefont {Burn},
			\citenamefont {Zhang}, \citenamefont {Yu}, \citenamefont {Guang},
			\citenamefont {Chen}, \citenamefont {Qiu}, \citenamefont {van~der Laan},\
			and\ \citenamefont {Hesjedal}}]{Burn2020}%
		\BibitemOpen
		\bibfield  {author} {\bibinfo {author} {\bibfnamefont {D.~M.}\ \bibnamefont
				{Burn}}, \bibinfo {author} {\bibfnamefont {S.~L.}\ \bibnamefont {Zhang}},
			\bibinfo {author} {\bibfnamefont {G.~Q.}\ \bibnamefont {Yu}}, \bibinfo
			{author} {\bibfnamefont {Y.}~\bibnamefont {Guang}}, \bibinfo {author}
			{\bibfnamefont {H.~J.}\ \bibnamefont {Chen}}, \bibinfo {author}
			{\bibfnamefont {X.~P.}\ \bibnamefont {Qiu}}, \bibinfo {author} {\bibfnamefont
				{G.}~\bibnamefont {van~der Laan}},\ and\ \bibinfo {author} {\bibfnamefont
				{T.}~\bibnamefont {Hesjedal}},\ }\bibfield  {title} {\bibinfo {title}
			{{Depth-Resolved Magnetization Dynamics Revealed by X-Ray Reflectometry
					Ferromagnetic Resonance}},\ }\href
		{https://doi.org/10.1103/PhysRevLett.125.137201} {\bibfield  {journal}
			{\bibinfo  {journal} {Phys. Rev. Lett.}\ }\textbf {\bibinfo {volume} {125}},\
			\bibinfo {pages} {137201} (\bibinfo {year} {2020})}\BibitemShut {NoStop}%
		\bibitem [{\citenamefont {Radu}\ \emph {et~al.}(2011)\citenamefont {Radu},
			\citenamefont {Vahaplar}, \citenamefont {Stamm}, \citenamefont {Kachel},
			\citenamefont {Pontius}, \citenamefont {D{\"u}rr}, \citenamefont {Ostler},
			\citenamefont {Barker}, \citenamefont {Evans}, \citenamefont {Chantrell},
			\citenamefont {Tsukamoto}, \citenamefont {Itoh}, \citenamefont {Kirilyuk},
			\citenamefont {Rasing},\ and\ \citenamefont {Kimel}}]{Radu2011}%
		\BibitemOpen
		\bibfield  {author} {\bibinfo {author} {\bibfnamefont {I.}~\bibnamefont
				{Radu}}, \bibinfo {author} {\bibfnamefont {K.}~\bibnamefont {Vahaplar}},
			\bibinfo {author} {\bibfnamefont {C.}~\bibnamefont {Stamm}}, \bibinfo
			{author} {\bibfnamefont {T.}~\bibnamefont {Kachel}}, \bibinfo {author}
			{\bibfnamefont {N.}~\bibnamefont {Pontius}}, \bibinfo {author} {\bibfnamefont
				{H.~A.}\ \bibnamefont {D{\"u}rr}}, \bibinfo {author} {\bibfnamefont {T.~A.}\
				\bibnamefont {Ostler}}, \bibinfo {author} {\bibfnamefont {J.}~\bibnamefont
				{Barker}}, \bibinfo {author} {\bibfnamefont {R.~F.~L.}\ \bibnamefont
				{Evans}}, \bibinfo {author} {\bibfnamefont {R.~W.}\ \bibnamefont
				{Chantrell}}, \bibinfo {author} {\bibfnamefont {A.}~\bibnamefont
				{Tsukamoto}}, \bibinfo {author} {\bibfnamefont {A.}~\bibnamefont {Itoh}},
			\bibinfo {author} {\bibfnamefont {A.}~\bibnamefont {Kirilyuk}}, \bibinfo
			{author} {\bibfnamefont {T.}~\bibnamefont {Rasing}},\ and\ \bibinfo {author}
			{\bibfnamefont {A.~V.}\ \bibnamefont {Kimel}},\ }\bibfield  {title} {\bibinfo
			{title} {Transient ferromagnetic-like state mediating ultrafast reversal of
				antiferromagnetically coupled spins},\ }\href
		{https://doi.org/10.1038/nature09901} {\bibfield  {journal} {\bibinfo
				{journal} {Nature}\ }\textbf {\bibinfo {volume} {472}},\ \bibinfo {pages}
			{205} (\bibinfo {year} {2011})}\BibitemShut {NoStop}%
		\bibitem [{\citenamefont {Steinbach}\ \emph {et~al.}(2022)\citenamefont
			{Steinbach}, \citenamefont {Stetzuhn}, \citenamefont {Engel}, \citenamefont
			{Atxitia}, \citenamefont {{von Korff Schmising}},\ and\ \citenamefont
			{Eisebitt}}]{Steinbach2022}%
		\BibitemOpen
		\bibfield  {author} {\bibinfo {author} {\bibfnamefont {F.}~\bibnamefont
				{Steinbach}}, \bibinfo {author} {\bibfnamefont {N.}~\bibnamefont {Stetzuhn}},
			\bibinfo {author} {\bibfnamefont {D.}~\bibnamefont {Engel}}, \bibinfo
			{author} {\bibfnamefont {U.}~\bibnamefont {Atxitia}}, \bibinfo {author}
			{\bibfnamefont {C.}~\bibnamefont {{von Korff Schmising}}},\ and\ \bibinfo
			{author} {\bibfnamefont {S.}~\bibnamefont {Eisebitt}},\ }\bibfield  {title}
		{\bibinfo {title} {Accelerating double pulse all-optical write/erase cycles
				in metallic ferrimagnets},\ }\href {https://doi.org/10.1063/5.0080351}
		{\bibfield  {journal} {\bibinfo  {journal} {Appl. Phys. Lett.}\ }\textbf
			{\bibinfo {volume} {120}},\ \bibinfo {pages} {112406} (\bibinfo {year}
			{2022})}\BibitemShut {NoStop}%
		\bibitem [{\citenamefont {Igarashi}\ \emph {et~al.}(2020)\citenamefont
			{Igarashi}, \citenamefont {Remy}, \citenamefont {Iihama}, \citenamefont
			{Malinowski}, \citenamefont {Hehn}, \citenamefont {Gorchon}, \citenamefont
			{Hohlfeld}, \citenamefont {Fukami}, \citenamefont {Ohno},\ and\ \citenamefont
			{Mangin}}]{igarashi2020}%
		\BibitemOpen
		\bibfield  {author} {\bibinfo {author} {\bibfnamefont {J.}~\bibnamefont
				{Igarashi}}, \bibinfo {author} {\bibfnamefont {Q.}~\bibnamefont {Remy}},
			\bibinfo {author} {\bibfnamefont {S.}~\bibnamefont {Iihama}}, \bibinfo
			{author} {\bibfnamefont {G.}~\bibnamefont {Malinowski}}, \bibinfo {author}
			{\bibfnamefont {M.}~\bibnamefont {Hehn}}, \bibinfo {author} {\bibfnamefont
				{J.}~\bibnamefont {Gorchon}}, \bibinfo {author} {\bibfnamefont
				{J.}~\bibnamefont {Hohlfeld}}, \bibinfo {author} {\bibfnamefont
				{S.}~\bibnamefont {Fukami}}, \bibinfo {author} {\bibfnamefont
				{H.}~\bibnamefont {Ohno}},\ and\ \bibinfo {author} {\bibfnamefont
				{S.}~\bibnamefont {Mangin}},\ }\bibfield  {title} {\bibinfo {title}
			{Engineering single-shot all-optical switching of ferromagnetic materials},\
		}\href {https://doi.org/10.1021/acs.nanolett.0c03373} {\bibfield  {journal}
			{\bibinfo  {journal} {Nano Lett.}\ }\textbf {\bibinfo {volume} {20}},\
			\bibinfo {pages} {8654} (\bibinfo {year} {2020})}\BibitemShut {NoStop}%
		\bibitem [{\citenamefont {Yang}\ \emph {et~al.}(2017)\citenamefont {Yang},
			\citenamefont {Wilson}, \citenamefont {Gorchon}, \citenamefont {Lambert},
			\citenamefont {Salahuddin},\ and\ \citenamefont {Bokor}}]{Yang2017}%
		\BibitemOpen
		\bibfield  {author} {\bibinfo {author} {\bibfnamefont {Y.}~\bibnamefont
				{Yang}}, \bibinfo {author} {\bibfnamefont {R.~B.}\ \bibnamefont {Wilson}},
			\bibinfo {author} {\bibfnamefont {J.}~\bibnamefont {Gorchon}}, \bibinfo
			{author} {\bibfnamefont {C.~H.}\ \bibnamefont {Lambert}}, \bibinfo {author}
			{\bibfnamefont {S.}~\bibnamefont {Salahuddin}},\ and\ \bibinfo {author}
			{\bibfnamefont {J.}~\bibnamefont {Bokor}},\ }\bibfield  {title} {\bibinfo
			{title} {{Ultrafast magnetization reversal by picosecond electrical
					pulses}},\ }\href {https://doi.org/10.1126/sciadv.1603117} {\bibfield
			{journal} {\bibinfo  {journal} {Sci. Adv.}\ }\textbf {\bibinfo {volume}
				{3}},\ \bibinfo {pages} {1} (\bibinfo {year} {2017})},\ \Eprint
		{https://arxiv.org/abs/1609.06392} {1609.06392} \BibitemShut {NoStop}%
		\bibitem [{\citenamefont {Grimaldi}\ \emph {et~al.}(2020)\citenamefont
			{Grimaldi}, \citenamefont {Krizakova}, \citenamefont {Sala}, \citenamefont
			{Yasin}, \citenamefont {Couet}, \citenamefont {Sankar~Kar}, \citenamefont
			{Garello},\ and\ \citenamefont {Gambardella}}]{Grimaldi2020}%
		\BibitemOpen
		\bibfield  {author} {\bibinfo {author} {\bibfnamefont {E.}~\bibnamefont
				{Grimaldi}}, \bibinfo {author} {\bibfnamefont {V.}~\bibnamefont {Krizakova}},
			\bibinfo {author} {\bibfnamefont {G.}~\bibnamefont {Sala}}, \bibinfo {author}
			{\bibfnamefont {F.}~\bibnamefont {Yasin}}, \bibinfo {author} {\bibfnamefont
				{S.}~\bibnamefont {Couet}}, \bibinfo {author} {\bibfnamefont
				{G.}~\bibnamefont {Sankar~Kar}}, \bibinfo {author} {\bibfnamefont
				{K.}~\bibnamefont {Garello}},\ and\ \bibinfo {author} {\bibfnamefont
				{P.}~\bibnamefont {Gambardella}},\ }\bibfield  {title} {\bibinfo {title}
			{Single-shot dynamics of spin\textendash orbit torque and spin transfer
				torque switching in three-terminal magnetic tunnel junctions},\ }\href
		{https://doi.org/10.1038/s41565-019-0607-7} {\bibfield  {journal} {\bibinfo
				{journal} {Nat. Nanotechnol.}\ }\textbf {\bibinfo {volume} {15}},\ \bibinfo
			{pages} {111} (\bibinfo {year} {2020})}\BibitemShut {NoStop}%
	\end{thebibliography}

\begin{thebibliography}{48}%
	\makeatletter
	\providecommand \@ifxundefined [1]{%
		\@ifx{#1\undefined}
	}%
	\providecommand \@ifnum [1]{%
		\ifnum #1\expandafter \@firstoftwo
		\else \expandafter \@secondoftwo
		\fi
	}%
	\providecommand \@ifx [1]{%
		\ifx #1\expandafter \@firstoftwo
		\else \expandafter \@secondoftwo
		\fi
	}%
	\providecommand \natexlab [1]{#1}%
	\providecommand \enquote  [1]{``#1''}%
	\providecommand \bibnamefont  [1]{#1}%
	\providecommand \bibfnamefont [1]{#1}%
	\providecommand \citenamefont [1]{#1}%
	\providecommand \href@noop [0]{\@secondoftwo}%
	\providecommand \href [0]{\begingroup \@sanitize@url \@href}%
	\providecommand \@href[1]{\@@startlink{#1}\@@href}%
	\providecommand \@@href[1]{\endgroup#1\@@endlink}%
	\providecommand \@sanitize@url [0]{\catcode `\\12\catcode `\$12\catcode
		`\&12\catcode `\#12\catcode `\^12\catcode `\_12\catcode `\%12\relax}%
	\providecommand \@@startlink[1]{}%
	\providecommand \@@endlink[0]{}%
	\providecommand \url  [0]{\begingroup\@sanitize@url \@url }%
	\providecommand \@url [1]{\endgroup\@href {#1}{\urlprefix }}%
	\providecommand \urlprefix  [0]{URL }%
	\providecommand \Eprint [0]{\href }%
	\providecommand \doibase [0]{https://doi.org/}%
	\providecommand \selectlanguage [0]{\@gobble}%
	\providecommand \bibinfo  [0]{\@secondoftwo}%
	\providecommand \bibfield  [0]{\@secondoftwo}%
	\providecommand \translation [1]{[#1]}%
	\providecommand \BibitemOpen [0]{}%
	\providecommand \bibitemStop [0]{}%
	\providecommand \bibitemNoStop [0]{.\EOS\space}%
	\providecommand \EOS [0]{\spacefactor3000\relax}%
	\providecommand \BibitemShut  [1]{\csname bibitem#1\endcsname}%
	\let\auto@bib@innerbib\@empty

	\bibitem [{\citenamefont {Sch{\"{u}}tz}\ \emph {et~al.}(1987)\citenamefont
		{Sch{\"{u}}tz}, \citenamefont {Wagner}, \citenamefont {Wilhelm},
		\citenamefont {Kienle}, \citenamefont {Zeller}, \citenamefont {Frahm},\ and\
		\citenamefont {Materlik}}]{Schutz1987}%
	\BibitemOpen
	\bibfield  {author} {\bibinfo {author} {\bibfnamefont {G.}~\bibnamefont
			{Sch{\"{u}}tz}}, \bibinfo {author} {\bibfnamefont {W.}~\bibnamefont
			{Wagner}}, \bibinfo {author} {\bibfnamefont {W.}~\bibnamefont {Wilhelm}},
		\bibinfo {author} {\bibfnamefont {P.}~\bibnamefont {Kienle}}, \bibinfo
		{author} {\bibfnamefont {R.}~\bibnamefont {Zeller}}, \bibinfo {author}
		{\bibfnamefont {R.}~\bibnamefont {Frahm}},\ and\ \bibinfo {author}
		{\bibfnamefont {G.}~\bibnamefont {Materlik}},\ }\bibfield  {title} {\bibinfo
		{title} {{Absorption of circularly polarized x rays in iron}},\ }\href
	{https://doi.org/10.1103/PhysRevLett.58.737} {\bibfield  {journal} {\bibinfo
			{journal} {Phys. Rev. Lett.}\ }\textbf {\bibinfo {volume} {58}},\ \bibinfo
		{pages} {737} (\bibinfo {year} {1987})}\BibitemShut {NoStop}%
	\bibitem [{\citenamefont {Schick}(2021)}]{Schick2021b}%
	\BibitemOpen
	\bibfield  {author} {\bibinfo {author} {\bibfnamefont {D.}~\bibnamefont
			{Schick}},\ }\bibfield  {title} {\bibinfo {title} {{udkm1Dsim – a Python
				toolbox for simulating 1D ultrafast dynamics in condensed matter}},\ }\href
	{https://doi.org/10.1016/j.cpc.2021.108031} {\bibfield  {journal} {\bibinfo
			{journal} {Comput. Phys. Commun.}\ }\textbf {\bibinfo {volume} {266}},\
		\bibinfo {pages} {108031} (\bibinfo {year} {2021})},\ \Eprint
	{https://arxiv.org/abs/2102.12144} {2102.12144} \BibitemShut {NoStop}%
	\bibitem [{\citenamefont {Chantler}\ and\ \citenamefont
		{IUCr}(2001)}]{Chantler2001}%
	\BibitemOpen
	\bibfield  {author} {\bibinfo {author} {\bibfnamefont {C.~T.}\ \bibnamefont
			{Chantler}}\ and\ \bibinfo {author} {\bibnamefont {IUCr}},\ }\href
	{https://doi.org/10.1107/S0909049501008305} {\bibinfo {title} {Detailed
			tabulation of atomic form factors, photoelectric absorption and scattering
			cross section, and mass attenuation coefficients in the vicinity of
			absorption edges in the soft x-ray (z = 30-36, z = 60-89, e = 0.1-10 kev) -
			addressing convergence issues of earlier work}} (\bibinfo {year}
	{2001})\BibitemShut {NoStop}%
	\bibitem [{\citenamefont {Chen}\ \emph {et~al.}(1995)\citenamefont {Chen},
		\citenamefont {Idzerda}, \citenamefont {Lin}, \citenamefont {Smith},
		\citenamefont {Meigs}, \citenamefont {Chaban}, \citenamefont {Ho},
		\citenamefont {Pellegrin},\ and\ \citenamefont {Sette}}]{Chen1995}%
	\BibitemOpen
	\bibfield  {author} {\bibinfo {author} {\bibfnamefont {C.~T.}\ \bibnamefont
			{Chen}}, \bibinfo {author} {\bibfnamefont {Y.~U.}\ \bibnamefont {Idzerda}},
		\bibinfo {author} {\bibfnamefont {H.-J.}\ \bibnamefont {Lin}}, \bibinfo
		{author} {\bibfnamefont {N.~V.}\ \bibnamefont {Smith}}, \bibinfo {author}
		{\bibfnamefont {G.}~\bibnamefont {Meigs}}, \bibinfo {author} {\bibfnamefont
			{E.}~\bibnamefont {Chaban}}, \bibinfo {author} {\bibfnamefont {G.~H.}\
			\bibnamefont {Ho}}, \bibinfo {author} {\bibfnamefont {E.}~\bibnamefont
			{Pellegrin}},\ and\ \bibinfo {author} {\bibfnamefont {F.}~\bibnamefont
			{Sette}},\ }\bibfield  {title} {\bibinfo {title} {{Experimental Confirmation
				of the X-Ray Magnetic Circular Dichroism Sum Rules for Iron and Cobalt}},\
	}\href {https://doi.org/10.1103/PhysRevLett.75.152} {\bibfield  {journal}
		{\bibinfo  {journal} {Phys. Rev. Lett.}\ }\textbf {\bibinfo {volume} {75}},\
		\bibinfo {pages} {152} (\bibinfo {year} {1995})}\BibitemShut {NoStop}%
	\bibitem [{\citenamefont {Peters}\ \emph {et~al.}(2004)\citenamefont {Peters},
		\citenamefont {Miguel}, \citenamefont {{de Vries}}, \citenamefont
		{Toulemonde}, \citenamefont {Goedkoop}, \citenamefont {Dhesi},\ and\
		\citenamefont {Brookes}}]{Peters2004}%
	\BibitemOpen
	\bibfield  {author} {\bibinfo {author} {\bibfnamefont {J.~F.}\ \bibnamefont
			{Peters}}, \bibinfo {author} {\bibfnamefont {J.}~\bibnamefont {Miguel}},
		\bibinfo {author} {\bibfnamefont {M.~A.}\ \bibnamefont {{de Vries}}},
		\bibinfo {author} {\bibfnamefont {O.~M.}\ \bibnamefont {Toulemonde}},
		\bibinfo {author} {\bibfnamefont {J.~B.}\ \bibnamefont {Goedkoop}}, \bibinfo
		{author} {\bibfnamefont {S.~S.}\ \bibnamefont {Dhesi}},\ and\ \bibinfo
		{author} {\bibfnamefont {N.~B.}\ \bibnamefont {Brookes}},\ }\bibfield
	{title} {\bibinfo {title} {Soft x-ray resonant magneto-optical constants at
			the gd ${M}_{4,5}$ and fe ${L}_{2,3}$ edges},\ }\href
	{https://doi.org/10.1103/PhysRevB.70.224417} {\bibfield  {journal} {\bibinfo
			{journal} {Phys. Rev. B}\ }\textbf {\bibinfo {volume} {70}},\ \bibinfo
		{pages} {224417} (\bibinfo {year} {2004})}\BibitemShut {NoStop}%
	\bibitem [{\citenamefont {Kachel}(2016)}]{kachel_pm3_2016}%
	\BibitemOpen
	\bibfield  {author} {\bibinfo {author} {\bibfnamefont {T.}~\bibnamefont
			{Kachel}},\ }\bibfield  {title} {\bibinfo {title} {The {PM3} beamline at
			{BESSY} {II}},\ }\href {https://doi.org/10.17815/jlsrf-2-73} {\bibfield
		{journal} {\bibinfo  {journal} {Journal of large-scale research facilities
				JLSRF}\ }\textbf {\bibinfo {volume} {2}},\ \bibinfo {pages} {48} (\bibinfo
		{year} {2016})},\ \bibinfo {note} {number: 0}\BibitemShut {NoStop}%
\end{thebibliography}

%

\end{document}